\RequirePackage{fix-cm}
\documentclass[12pt]{iopjournal}

\usepackage{amsmath,amssymb,amsthm}
\usepackage{textcase}
\usepackage[shortlabels]{enumitem}
\usepackage{array}
\usepackage{booktabs}
\usepackage{orcidlink}
\usepackage{etoolbox}
\usepackage[left=20mm,right=20mm,top=18mm,bottom=24mm,footskip=10mm]{geometry}
\usepackage[expansion=false]{microtype}
\patchcmd{\abstract}{\fontsize{11}{12}}{\fontsize{12}{14.5}}{}{}
\renewcommand{\articletype}[1]{\noindent\textsf{#1}\par\medskip}
\setlist{topsep=4pt,itemsep=2pt,parsep=0pt,partopsep=0pt}
\AtBeginDocument{%
  \setlength{\abovedisplayskip}{7pt plus 2pt minus 2pt}%
  \setlength{\belowdisplayskip}{7pt plus 2pt minus 2pt}%
  \setlength{\abovedisplayshortskip}{3pt plus 2pt}%
  \setlength{\belowdisplayshortskip}{5pt plus 2pt minus 2pt}%
}
\apptocmd{\thebibliography}{\setlength{\itemsep}{0pt}\setlength{\parsep}{0pt}}{}{}
\renewcommand{\orcid}[1]{\orcidlink{#1}}
\graphicspath{{./}}
\newcolumntype{L}[1]{>{\raggedright\arraybackslash}p{#1}}

\newcommand{\figIfExists}[2][width=0.75\textwidth]{%
  \includegraphics[#1]{#2}%
}

\newcommand{\cH}{\mathcal{H}}
\newcommand{\cB}{\mathcal{B}}
\newcommand{\cU}{\mathrm{U}}
\newcommand{\cS}{\mathcal{S}}

\newcommand{\ket}[1]{\ensuremath{\left|#1\right\rangle}}
\newcommand{\bra}[1]{\ensuremath{\left\langle#1\right|}}
\newcommand{\rhoH}{\rho_{\mathcal{S}}}
\newcommand{\Span}{\operatorname{span}}
\newcommand{\tr}{\operatorname{tr}}
\newcommand{\Hsep}{\mathbf{H}_{\mathrm{sep}}}
\newcommand{\Hns}{\mathbf{H}_{\mathrm{ns}}}
\newcommand{\op}{\mathrm{op}}
\newcommand{\HS}{\mathrm{HS}}
\newcommand{\Pauli}{\mathcal{P}}

\newtheorem{theorem}{Theorem}[section]
\newtheorem{corollary}[theorem]{Corollary}
\newtheorem{lemma}[theorem]{Lemma}
\newtheorem{proposition}[theorem]{Proposition}
\theoremstyle{definition}

\theoremstyle{remark}
\newtheorem{remark}[theorem]{Remark}
\newtheorem*{example}{Example}
\numberwithin{equation}{section}

\begin{document}

\articletype{Paper}

\title{A Hamiltonian-Level Certificate for Network-Free Distributed Quantum Simulation:
Exact Tensor-Separability Criterion and Approximate Residual Bounds}

\author{Kan~He$^{1,\dagger}$\orcid{0000-0002-4931-2115}, Shusen~Liu$^{3,4,*,\dagger}$\orcid{0000-0002-0794-1132}, Jinchuan~Hou$^{2}$\orcid{0000-0001-6534-9228}, Pascal Jahan Elahi$^{3,4}$\orcid{0000-0002-6154-7224}, Yifei~Huang$^{2}$\orcid{0009-0007-0198-8734},
Xinyuan~Yu$^{5}$\orcid{0009-0009-4024-3668} and Ugo Varetto$^{3,4}$\orcid{0000-0002-7696-0345}}

\affil{$^1$College of Information and Computer and College of Mathematics, Taiyuan University of Technology, Taiyuan 030024, China}

\affil{$^2$College of Mathematics, Taiyuan University of Technology, Taiyuan 030024, China}

\affil{$^3$Pawsey Supercomputing Research Centre, Kensington, WA 6151, Australia}

\affil{$^4$School of Physics, Mathematics and Computing, The University of Western Australia, Perth, WA 6009, Australia}

\affil{$^5$Independent Researcher, Guangzhou 510000, China}

\affil{$^\dagger$K.~He and S.~Liu contributed equally to this work.}

\affil{$^*$Author to whom any correspondence should be addressed.}

\email{shusen.liu@csiro.au}


\begin{abstract}
Circuit cutting allows quantum circuits to be evaluated on smaller devices at the cost of additional sampling and classical reconstruction. For Hamiltonian-simulation circuits, the target unitary $\exp[it\mathbf H]$ connects this decomposition problem to the structure of the generator. We develop a Hamiltonian-level certificate that assesses exact separability and the accuracy of independent module evolution before a gate decomposition is chosen. For a time-independent Hamiltonian and a fixed module partition, the Hilbert--Schmidt projection onto sums of module-local terms isolates a cross-module residual $\Hns$. Its vanishing is necessary and sufficient for product evolution at all times; for Pauli-list inputs, this condition is checked in one pass over the combined coefficients. When $\Hns\ne0$, the projected evolution has operator-norm error at most $|t|\,\|\Hns\|_\op$. Across a bipartition, an explicit first-order algebraic correction has the same operator-Schmidt rank as $\Hns$ and leaves a second-order residual. The same Hamiltonian residual connects this approximation to short-time entanglement, determining the leading state-space and operator-space entangling powers and, through its Schmidt spectrum, the leading Choi tripartite-information terms. Numerical experiments test these relations and demonstrate block-factorised simulation up to $20$ qubits.
\end{abstract}

\section{Introduction}\label{sec:intro}

\subsection{Distributed quantum computing without a quantum network}\label{sec:lead-in}

Consider a $50$-qubit Hamiltonian simulation assigned to two $25$-qubit modules. Each module can run local circuits, but the modules have no quantum channel or shared entanglement. Which evolutions can they implement independently, and what error results from omitting the interactions between them? This is the network-free limit of distributed architectures that use entanglement and teleportation to connect smaller processors~\cite{Cirac1999,Sundaram2023}. We study a fixed assignment of qubits to modules and a supplied, time-independent Hamiltonian.

The answer is determined by the Hilbert--Schmidt projection $\Hsep$ of $\mathbf{H}$ onto the block-additive Hamiltonians, which are sums of terms acting within individual modules. Its residual $\Hns=\mathbf{H}-\Hsep$ contains the cross-module interactions. The evolution $\exp[it\mathbf{H}]$ factors across the partition for every $t$ if and only if $\Hns=0$, equivalently $\rhoH=0$, where $\rhoH$ is the fraction of squared Hilbert--Schmidt norm carried by the residual. We call this test the \emph{network-free certificate}. It concerns the target unitary itself; preparing its input state and reconstructing global measurement outputs are separate tasks.

Away from this boundary, independent evolution under $\Hsep$ has operator-norm error at most $|t|\,\|\Hns\|_\op$. For a two-module split, an explicit first-order correction has the same operator-Schmidt rank as $\Hns$ and leaves a second-order residual. These results give a time-dependent error guarantee and the product-term structure of a possible correction; realising that correction requires a specified protocol. Quantitative guarantees depend on $\Hns$ itself: its norms are invariant under scalar shifts of $\mathbf{H}$, whereas the denominator of $\rhoH$ depends on the energy offset.

\subsection{Relation to cutting and scrambling}\label{sec:lead-in-context}

Circuit cutting and knitting address circuits that exceed a device's capacity by expressing global quantities through classically recombined local experiments. This approach builds on classical-simulation trade-offs~\cite{Bravyi2016}, wire cutting~\cite{Peng2020}, gate-cutting decompositions~\cite{Mitarai2021,Mitarai2021a}, communication-assisted knitting~\cite{Piveteau2023,Brenner2025}, and automated cut placement~\cite{Tang2021}. Our certificate can be evaluated before these protocols are chosen: it tests whether a Hamiltonian evolution already factors across the supplied partition and bounds the error of retaining only its block-additive part. Gate-level cutting costs are compared in \S\ref{ssec:auxiliary}, and related approaches in \S\ref{sec:related}.

For a qubit Hamiltonian supplied as a Pauli list, the projection retains precisely the strings supported within one module. After identical strings are combined, one pass over the coefficients identifies $\Hns$ and computes $\rhoH$. The input is therefore different from unitary-access methods that infer tensor-product structure from Choi-state data~\cite{Mansuroglu2024}, or the canonical (Cartan) decomposition of a known two-qubit gate~\cite{Kraus2001}. Quantum mereology instead treats the tensor-product structure itself as a variable~\cite{Soulas2025,Zanardi2026}; here it is fixed by the hardware or compiler assignment.

The all-time requirement is essential. A cross-module interaction can generate a product unitary at isolated times, as in $\exp[i\pi Z\otimes Z]=-I$; for bipartite single-tensor generators we give the exact resonance condition. The residual also connects the separability criterion to short-time dynamics: across a bipartition, $\|\Hns\|_\HS^2$ determines the leading state-space and operator-space entangling powers and enters the leading Choi-information expansion. In particular, it is the squared residual norm appearing in the Gaussian scrambling rate of Andreadakis--Dallas--Zanardi (ADZ)~\cite{Andreadakis2024}.

The main technical statements are as follows:
\begin{itemize}
\item Theorem~\ref{thm:pauli-schmidt} characterises all-time product evolution for any partition of finite-dimensional factors. Theorem~\ref{thm:bipartite-single-t} characterises single-time resonances for bipartite single-tensor generators.
\item Theorem~\ref{thm:approx} and Proposition~\ref{prop:HS} bound the block-parallel error in operator and Hilbert--Schmidt norms. Theorem~\ref{thm:schmidt-refinement} gives an explicit first-order correction, its operator-Schmidt rank, and a second-order residual bound.
\item Propositions~\ref{prop:ep-exact}, \ref{prop:rho-Ep}, and~\ref{prop:choi-exact} relate $\Hns$ to short-time entangling powers and Choi tripartite information. Numerical experiments test these relations and the approximation bounds, and demonstrate block-factorised simulation up to $20$ qubits.
\end{itemize}

The rest of the paper is organised as follows. \S\ref{sec:prelims} fixes notation, defines $\Hns$ and $\rhoH$, and proves the exact separability criterion. \S\ref{sec:exact} reads the criterion off the usual tensor presentations and proves the single-time resonance theorem. \S\ref{sec:approx} establishes the approximation and rank-refinement bounds. \S\ref{sec:experiments} reports the numerical experiments. \S\ref{sec:related} discusses related work, and \S\ref{sec:conclusion} concludes.

\section{Preliminaries and the exact criterion}\label{sec:prelims}

\paragraph{Standing assumption.} Sections~\ref{sec:prelims}--\ref{sec:experiments} are stated in finite dimensions unless a separate extension is explicitly noted, and all dynamics considered are the closed-system unitary flows $\exp[it\mathbf{H}]$ of a time-independent Hamiltonian. Infinite-dimensional variants of the exact tensor arguments require slice-map hypotheses (Remark~\ref{rem:bipartite-traceclass}) and are not used later.

\paragraph{Notation.} Let $\cH_1,\dots,\cH_n$ be finite-dimensional complex Hilbert spaces and $\cH:=\bigotimes_{k=1}^{n}\cH_k$ their tensor product. Write $\cB(\cH)$, $\cB_s(\cH)$, $\cU(\cH)$ for, respectively, the bounded operators, bounded self-adjoint operators, and unitary operators on $\cH$, and set $d_k:=\dim\cH_k$ and $d:=\dim\cH=\prod_{k}d_k$; for a subset $S\subseteq\{1,\dots,n\}$ write $\cH_S:=\bigotimes_{j\in S}\cH_j$ and $d_S:=\dim\cH_S=\prod_{j\in S}d_j$. $I$ denotes the identity, and $I_S$ (also $I_k$ for $S=\{k\}$, and $I_A$, $I_B$ for named blocks) the identity on $\cH_S$; spans of Pauli strings taken inside $\cB_s(\cH)$ are real spans. The partial trace $\tr_S$ traces out the factor $\cH_S$; explicitly $\tr_2(M\otimes\ket{y}\bra{y'})=\langle y'|y\rangle\, M$ on $\cH_1\otimes\cH_2$ for $M\in\cB(\cH_1)$ and $\ket{y},\ket{y'}\in\cH_2$, extended by linearity. For $A\in\cB(\cH)$, $\|A\|_\op=\sup_{\|\psi\|=1}\|A\psi\|$ denotes the operator norm (largest singular value) and $\|A\|_\HS=\sqrt{\tr(A^\dagger A)}$ the unnormalised Hilbert--Schmidt norm, with associated inner product $\langle M,N\rangle_\HS=\tr(M^\dagger N)$. A normalised convention $d^{-1}\tr(M^\dagger N)$ is used only where explicitly noted (the normalised mass $d^{-1}\|\mathbf{H}\|_\HS^2=\sum_Pc_P^2$ below and in the Choi-information fits of \S\ref{ssec:auxiliary}; the entangling-power results of \S\ref{ssec:connections} state their unnormalised convention explicitly); the ratio $\rhoH$ is unchanged by the choice.

\paragraph{Partitions.} A \emph{partition} $\cS=\{S_1,\dots,S_m\}$ of $\{1,\dots,n\}$ groups the tensor factors into $m$ blocks $\cH_{S_k}=\bigotimes_{j\in S_k}\cH_j$. In the distributed setting each block $S_k$ is the set of qubits held by one hardware module; for a bipartition we also call the two blocks the two sides of the cut. A unitary $U\in\cU(\cH)$ is \emph{$\cS$-separable} (or simply \emph{separable} when $\cS$ is clear) if $U=U_1\otimes\cdots\otimes U_m$ with $U_k\in\cU(\cH_{S_k})$ up to a global phase. A self-adjoint $\mathbf{H}\in\cB_s(\cH)$ is \emph{$\cS$-additive} if
\begin{equation}\label{eq:S-additive}
\mathbf{H} = \sum_{k=1}^{m} I^{\otimes \overline{S_k}}\otimes H_k,\qquad H_k\in\cB_s(\cH_{S_k}).
\end{equation}
Here $\overline{S_k}:=\{1,\dots,n\}\setminus S_k$ denotes the complement block, and the identity tensor is placed on those factors. We also call such $\mathbf{H}$ \emph{block-additive}, and the Hilbert--Schmidt projection onto the $\cS$-additive subspace the \emph{block-additive} projection; the two adjectives are used interchangeably for $\mathbf{H}$. (We reserve \emph{block-local} for an operator supported on a single block.)
If $\mathbf{H}$ is $\cS$-additive then its block summands commute, so
\[
\exp[it\mathbf{H}]
=\prod_{k=1}^{m}\exp\!\big[it(I^{\otimes\overline{S_k}}\otimes H_k)\big]
=\bigotimes_{k=1}^{m}\exp[itH_k],
\]
up to the fixed tensor-order identification of the blocks. Hence $\exp[it\mathbf{H}]$ is $\cS$-separable for every $t\in\mathbb{R}$. The converse is the content of Theorem~\ref{thm:pauli-schmidt} below.

\paragraph{Block-weight decomposition and the residual.} Write $\cB_0(\cH_S)$ for the traceless self-adjoint operators on $\cH_S$, embedded in $\cB_s(\cH)$ as $I^{\otimes\overline S}\otimes\cB_0(\cH_S)$ where needed. For each factor choose a Hilbert--Schmidt orthogonal basis of $\cB_s(\cH_k)$ consisting of $E^{(k)}_0:=I_k$ and a basis $\{E^{(k)}_\mu\}_{\mu\ge1}$ of $\cB_0(\cH_k)$. Their tensor products $E=E^{(1)}_{\mu_1}\otimes\cdots\otimes E^{(n)}_{\mu_n}$ form an orthogonal basis of $\cB_s(\cH)$ over $\mathbb{R}$ (traces of tensor products factorise, and there are $\prod_kd_k^2=d^2$ of them). Given a partition $\cS$, the \emph{block-weight} of a basis element is the number of blocks on which it acts non-trivially,
\begin{equation}\label{eq:block-weight}
w_\cS(E) = \#\{k\in\{1,\dots,m\}:\exists j\in S_k,\ E^{(j)}_{\mu_j}\neq I_j\}.
\end{equation}
Here $\#T$ denotes the cardinality of a finite set $T$. The block-weight sectors give the orthogonal decomposition $\cB_s(\cH)=\bigoplus_{w=0}^{m}\cB_s^{(w)}$ with $\cB_s^{(w)}:=\Span\{E:w_\cS(E)=w\}$. The sectors do not depend on the chosen local bases: $\cB_s^{(w)}=\bigoplus_{|K|=w}\bigotimes_{k\in K}\cB_0(\cH_{S_k})$ (tensored with the identity on the other blocks), since within one block the products with at least one traceless factor span the orthogonal complement of $I_{S_k}$, which is $\cB_0(\cH_{S_k})$. In particular $\cB_s^{(0)}=\mathbb{R}I$ and $\cB_s^{(1)}=\bigoplus_k\cB_{0}(\cH_{S_k})$, so that $\cB_s^{(0)}\oplus\cB_s^{(1)}$ is exactly the $\cS$-additive subspace (each block summand $H_k$ of~\eqref{eq:S-additive} splits as $\mathbb{R}I_{S_k}\oplus\cB_0(\cH_{S_k})$). Let $\Hsep$ be the Hilbert--Schmidt orthogonal projection of $\mathbf{H}$ onto the $\cS$-additive subspace and $\Hns:=\mathbf{H}-\Hsep\in\bigoplus_{w\ge2}\cB_s^{(w)}$ the \emph{non-separable residual}; we write $\Pi_{\mathrm{ns}}$ for the projection $\mathbf{H}\mapsto\Hns$. Writing $\mathcal E_k(X):=I^{\otimes\overline{S_k}}\otimes\big(\tr_{\overline{S_k}}X\big)/d_{\overline{S_k}}$ for the conditional expectation onto operators supported on block $k$ (the Hilbert--Schmidt projection onto $I^{\otimes\overline{S_k}}\otimes\cB(\cH_{S_k})$) and $\tau(X):=d^{-1}\tr(X)\,I$ for the projection onto $\mathbb{R}I$, the projection has the closed form
\begin{equation}\label{eq:Hsep-closed}
\Hsep=\sum_{k=1}^{m}\mathcal E_k(\mathbf{H})-(m-1)\,\tau(\mathbf{H}),\qquad
\Hns=\mathbf{H}-\sum_{k=1}^{m}\mathcal E_k(\mathbf{H})+(m-1)\,\tau(\mathbf{H}),
\end{equation}
since $\mathcal E_k=\tau+\Pi_k$ with $\Pi_k$ the projection onto $\cB_0(\cH_{S_k})$, and these sectors are mutually orthogonal and orthogonal to $I$. In the qubit case $\cH_k=\mathbb{C}^2$ the natural local basis is the Pauli set $\Pauli=\{I,X,Y,Z\}$ (also written $\sigma_X,\sigma_Y,\sigma_Z$ where that notation is clearer), the basis elements are the Pauli strings $P\in\Pauli^n$, and every $\mathbf{H}\in\cB_s((\mathbb{C}^2)^{\otimes n})$ has the unique expansion
\begin{equation}\label{eq:pauli-exp}
\mathbf{H}=\sum_{P\in\Pauli^n} c_P P,\qquad c_P\in\mathbb{R},\ c_P=2^{-n}\tr(P\,\mathbf{H}),
\end{equation}
with $\Hsep=\sum_{w_\cS(P)\le1}c_PP$ and $\Hns=\sum_{w_\cS(P)\ge2}c_PP$.

\paragraph{Non-separable mass.} We call $\|\Hns\|_\HS^2$ the \emph{non-separable mass} of $\mathbf{H}$. The scalar
\begin{equation}\label{eq:rhoH}
\rhoH(\mathbf{H}) := \frac{\|\Hns\|_\HS^2}{\|\mathbf{H}\|_\HS^2}\in[0,1]\qquad(\mathbf{H}\ne0)
\end{equation}
is the corresponding \emph{non-separable mass fraction}, the fraction of Hilbert--Schmidt mass lying outside the $\cS$-additive subspace; set $\rhoH(0):=0$. In the qubit Pauli basis (orthogonal, with $\|P\|_\HS^2=2^n$ for every string, so the factor $2^n$ cancels)
\[
\rhoH(\mathbf{H})
=\frac{\sum_{P:\,w_\cS(P)\ge 2} c_P^2}{\sum_{P} c_P^2}
\qquad(\mathbf{H}\ne0),
\]
so $\rhoH$ is exactly the fraction of squared Pauli coefficient weight on strings that touch at least two blocks, and $d^{-1}\|\Hns\|_\HS^2=\sum_{w_\cS(P)\ge2}c_P^2$ is the \emph{normalised} mass.

\begin{theorem}[Exact separability criterion]\label{thm:pauli-schmidt}
Let $\cS=\{S_1,\dots,S_m\}$ be a partition of $\{1,\dots,n\}$ with finite-dimensional factors and $\mathbf{H}\in\cB_s(\cH)$. The following are equivalent.
\begin{enumerate}[\rm (i)]
\item $\mathbf{H}$ is $\cS$-additive.
\item For every $t\in\mathbb{R}$, $\exp[it\mathbf{H}]$ is $\cS$-separable.
\item $\Hns=0$; equivalently $\mathbf{H}=\Hsep$, equivalently $\rhoH(\mathbf{H})=0$.
\item In any product basis of local orthogonal bases containing the identities, every coefficient of $\mathbf{H}$ on a basis element of block-weight $w_\cS\ge2$ vanishes; for qubit registers, $c_P=0$ in~\eqref{eq:pauli-exp} for every Pauli string $P$ with $w_\cS(P)\ge2$.
\end{enumerate}
\end{theorem}

\noindent For an $n$-qubit Hamiltonian the certificate is a single pass over the (typically sparse) nonzero Pauli coefficients of $\mathbf{H}$: the dynamics factor across the hardware partition $\cS$ for every time $t$ if and only if no Pauli string that touches two or more blocks carries a nonzero coefficient. It is independent of any chosen tensor-sum presentation of $\mathbf{H}$, and for general local dimensions it is the same statement in any local traceless basis.

\begin{proof}
(i)$\Rightarrow$(ii) follows by blockwise exponentiation. For the converse, the tensor-product homomorphism
\[
\Phi:\prod_{k=1}^{m}\cU(\cH_{S_k})\longrightarrow\cU(\cH),\qquad
(U_1,\dots,U_m)\longmapsto U_1\otimes\cdots\otimes U_m
\]
has compact, hence closed, image $G$, the group of $\cS$-separable unitaries. By the closed-subgroup and Lie-group homomorphism theorems~\cite{Hall2015}, its Lie algebra is the image of the differential of $\Phi$:
\[
\mathrm{Lie}(G)=\left\{i\sum_k I^{\otimes\overline{S_k}}\otimes H_k:\ H_k\in\cB_s(\cH_{S_k})\right\}.
\]
If $\exp[it\mathbf{H}]\in G$ for every $t$, its tangent at $t=0$ is $i\mathbf{H}\in\mathrm{Lie}(G)$, so $\mathbf{H}$ is $\cS$-additive. Finally, (i), (iii), and (iv) are equivalent because the additive subspace is $\cB_s^{(0)}\oplus\cB_s^{(1)}$: its orthogonal complement is spanned by the basis elements with block-weight at least two. The equivalence $\Hns=0\Leftrightarrow\rhoH=0$ includes $\mathbf{H}=0$ by convention.
\end{proof}

\begin{remark}[Scalar shifts]\label{rem:shift}
$\rhoH$ is a property of the presented Hamiltonian rather than of the unitary channel alone. A scalar shift $\mathbf{H}\mapsto\mathbf{H}+\alpha I$ changes the flow only by a global phase and leaves $\Hns$ unchanged (hence also the numerator of~\eqref{eq:rhoH}, the zero set $\rhoH=0$, and the bounds of \S\ref{sec:approx} that involve $\mathbf{H}$ only through $\Hns$, such as~\eqref{eq:approx-op} and~\eqref{eq:approx-HS}), but it changes the denominator, $\|\mathbf{H}+\alpha I\|_\HS^2=\|\mathbf{H}-d^{-1}\tr(\mathbf{H})\,I\|_\HS^2+d\,(\alpha+d^{-1}\tr\mathbf{H})^2$, which is minimised on the traceless part, so a large identity offset dilutes the fraction $\rhoH$. Where a shift-invariant fraction is preferred, evaluate $\rhoH$ on the traceless part $\mathbf{H}-d^{-1}\tr(\mathbf{H})\,I$: the value for the quantum approximate optimisation algorithm (QAOA) MaxCut cost operator in \S\ref{ssec:exp-boundary}, for instance, is $1/(|E|+1)$, with $|E|\ge1$ the edge count of the graph, when the identity offset is retained, and $1$ without it. For the same reason the leading bounds of \S\ref{sec:approx} are phrased in terms of the shift-invariant residual $\Hns$ (its operator and Hilbert--Schmidt norms) rather than the fraction $\rhoH$, whose role is the zero test of Theorem~\ref{thm:pauli-schmidt}.
\end{remark}

\paragraph{Two-qubit examples.} For $\theta\ne0$, the pure $XX$ coupling $\theta\, X\otimes X$ (the Mølmer--Sørensen interaction of trapped ions~\cite{Molmer1999}) has Pauli expansion concentrated on $XX$ (block-weight 2), so $\rhoH=1$: all its Hilbert--Schmidt mass lies in the residual. The single-site field $X\otimes I+I\otimes X$ has block-weight 1 only, so $\rhoH=0$. The controlled-NOT (CNOT) generator $\frac{\pi}{4}(I\otimes I-I\otimes X-Z\otimes I+Z\otimes X)$ has one cross-block term, $ZX$, among four equal-weight Pauli terms, giving $\rhoH=1/4$.

\section{Exact separability of tensor presentations}\label{sec:exact}

We now apply Theorem~\ref{thm:pauli-schmidt} to single tensors and sums of tensors, then characterise the exceptional times at which a bipartite single-tensor interaction generates a product unitary. Throughout, we abbreviate $\{\{1\},\dots,\{n\}\}$-additive and $\{\{1\},\dots,\{n\}\}$-separable as \emph{fully additive} and \emph{fully (tensor-)separable}.

\subsection{Single tensors}\label{ssec:bipartite}

For a self-adjoint $A$ on a finite-dimensional space $V$, write $A=A_0+\bar aI$ with $\bar a:=\tr(A)/\dim V$ and $A_0$ traceless.

\begin{corollary}[Bipartite single tensor]\label{thm:bipartite}
Let $\cH_1$ and $\cH_2$ be finite-dimensional, and let $A\in\cB_s(\cH_1)$ and $B\in\cB_s(\cH_2)$. The following are equivalent.
\begin{enumerate}[\rm (i)]
\item $A\otimes B$ is block-additive: there exist $H_1\in\cB_s(\cH_1)$, $H_2\in\cB_s(\cH_2)$ with $A\otimes B = H_1\otimes I + I\otimes H_2$. (This asks for a \emph{sum} of single-side generators, which a product operator need not be.)
\item For every $t\in\mathbb{R}$, $\exp[itA\otimes B]$ is tensor-separable: there exist unitaries $C_t\in\cU(\cH_1)$, $D_t\in\cU(\cH_2)$ with $\exp[itA\otimes B]=C_t\otimes D_t$ (any global phase can be absorbed into one local factor).
\item At least one of $A,B$ is a scalar multiple of the identity, $A\in\mathbb{R}I$ or $B\in\mathbb{R}I$.
\end{enumerate}
Moreover, if $B=\beta I$ one may take $C_t=\exp[it\beta A]$ and $D_t=I$ (and symmetrically when $A=\alpha I$).
\end{corollary}

\begin{proof}
(i)$\Leftrightarrow$(ii) is Theorem~\ref{thm:pauli-schmidt} for the bipartition. For (i)$\Leftrightarrow$(iii), expand $A\otimes B=A_0\otimes B_0+\bar b\,A_0\otimes I+\bar a\,I\otimes B_0+\bar a\bar b\,I$. The last three terms are $\cS$-additive and $A_0\otimes B_0$ lies in the block-weight-$2$ sector, so $\Pi_{\mathrm{ns}}(A\otimes B)=A_0\otimes B_0$, which vanishes if and only if $A_0=0$ or $B_0=0$, that is, if and only if $A$ or $B$ is scalar. The explicit factors in the scalar case are $\exp[itA\otimes\beta I]=\exp[it\beta A]\otimes I$.
\end{proof}

\begin{remark}[Uniqueness of the local factors]\label{rem:local-factor-uniqueness}
For each fixed $t$ the local factors in (ii) are unique up to a $\mathrm{U}(1)$ rebalance $C_t\mapsto e^{i\vartheta_t}C_t$, $D_t\mapsto e^{-i\vartheta_t}D_t$: if $C\otimes D=C'\otimes D'$ with all four factors unitary then $(C'^{-1}C)\otimes(D'^{-1}D)=I$, and the slice-map fact recorded in the proof of Theorem~\ref{thm:bipartite-single-t} ($M\otimes N=\lambda I$ with $\lambda\ne0$ forces both factors scalar) gives $C=e^{i\vartheta}C'$, $D=e^{-i\vartheta}D'$.
\end{remark}

\noindent Thus a single-tensor Hamiltonian generates product dynamics for all times only when it reduces to a single-side term. If both factors are non-scalar, product evolution is restricted to a discrete set of resonant times (Theorem~\ref{thm:bipartite-single-t}).

\begin{example}
On $\mathbb{C}^2\otimes\mathbb{C}^2$, take $A=B=\sigma_Z$, a two-sided interaction. Condition (i) fails: $\sigma_Z\otimes\sigma_Z$ is itself traceless$\,\otimes\,$traceless, so $\Pi_{\mathrm{ns}}(\sigma_Z\otimes\sigma_Z)=\sigma_Z\otimes\sigma_Z\ne0$ and $\rhoH=1$. Condition (ii) fails visibly: $\exp[it\,\sigma_Z\otimes\sigma_Z]$ maps the product state $\ket{+}\otimes\ket{+}$ to $\cos t\,\ket{+}\otimes\ket{+}+i\sin t\,\ket{-}\otimes\ket{-}$, which is entangled for every $t\notin\tfrac{\pi}{2}\mathbb{Z}$, whereas a product unitary $C_t\otimes D_t$ sends product states to product states. (The exceptional times $t\in\tfrac{\pi}{2}\mathbb{Z}$ are exactly the resonances of Theorem~\ref{thm:bipartite-single-t} below.) By contrast, the single-side term $\sigma_Z\otimes I$ satisfies all three conditions, with $\exp[it\,\sigma_Z\otimes I]=\exp[it\sigma_Z]\otimes I$ for every $t$.
\end{example}

\begin{remark}\label{rem:bipartite-traceclass}
The criterion is used only in finite dimensions. In infinite-dimensional variants, both the Hilbert--Schmidt decomposition and the compact-Lie-group argument behind Theorem~\ref{thm:pauli-schmidt} are unavailable, and the algebraic step (i)$\Rightarrow$(iii) of Corollary~\ref{thm:bipartite} would be replaced by bounded slice maps or vector-state functionals applied to $A\otimes B=H_1\otimes I+I\otimes H_2$; the Hilbert--Schmidt and entropy statements later in the paper are finite-dimensional only (Standing assumption) and are not extended.
\end{remark}

\begin{corollary}[Multipartite single tensor]\label{thm:multi-single}
Let $A_k\in\cB_s(\cH_k)$ for $k=1,\dots,n$, and set $\mathbf{H}=A_1\otimes A_2\otimes\cdots\otimes A_n$. The following are equivalent.
\begin{enumerate}[\rm (i)]
\item $\mathbf{H}$ is fully additive.
\item For every $t\in\mathbb{R}$, $\exp[it\mathbf{H}]$ is fully tensor-separable.
\item Either some $A_k=0$, or at most one element of $\{A_1,\dots,A_n\}$ fails to be a scalar multiple of the identity.
\end{enumerate}
\end{corollary}

\begin{proof}
(i)$\Leftrightarrow$(ii) is Theorem~\ref{thm:pauli-schmidt} for the singleton partition. If some $A_k=0$ then $\mathbf{H}=0$, and if at most one factor is non-scalar then $\mathbf{H}$ is a scalar multiple of a block-local operator; in both cases (i) holds. Conversely, assume every $A_k\ne0$ and let $T:=\{k:A_k\notin\mathbb{R}I\}$ have at least two elements. Expanding each factor as $A_k=A_{k,0}+\bar a_kI_k$, the term $\bigotimes_{k\in T}A_{k,0}\otimes\bigotimes_{k\notin T}\bar a_kI_k$ of the expansion of $\mathbf{H}$ is nonzero (each $A_{k,0}\ne0$ for $k\in T$, and $\bar a_k\ne0$ for $k\notin T$ since those factors are nonzero scalars), lies in the block-weight-$|T|$ sector with $|T|\ge2$, and is orthogonal to every other term of the expansion, so $\Hns\ne0$ and (i) fails by Theorem~\ref{thm:pauli-schmidt}(iii).
\end{proof}

\noindent If the tensor is nonzero and at least two factors are non-scalar, coarsen to a bipartition separating one such factor from the rest. Both grouped factors are non-scalar, so Theorem~\ref{thm:bipartite-single-t} restricts the fully separable times to a discrete set.

\begin{remark}\label{rem:formula}
In the nonzero case the flow is explicit: if at most one factor, say $A_j$, is non-scalar (any $j$ if all are scalar) and $A_k=\lambda_k I$ for $k\ne j$, set $\delta_j=\prod_{k\ne j}\lambda_k$ (an empty product is $1$). Then for every $t\in\mathbb{R}$,
\[ \exp[it\mathbf{H}] = I_1\otimes\cdots\otimes I_{j-1}\otimes \exp[it\delta_j A_j]\otimes I_{j+1}\otimes\cdots\otimes I_n. \]
The degenerate alternative $0\otimes\text{anything}=0$ is the zero Hamiltonian, which is additive and generates the identity flow regardless of the other factors.
\end{remark}

\begin{remark}[Sums of tensors]\label{thm:multi-sum}
Hamiltonians are usually presented as sums $\mathbf{H}=\sum_{i=1}^{N}T_i$, $T_i=A_i^{(1)}\otimes\cdots\otimes A_i^{(n)}$. If every nonzero displayed tensor has at most one non-scalar factor, then each $T_i$ is block-local and $\mathbf{H}$ is fully additive; commutativity of the summands is not needed for this. If in addition $[T_i,T_j]=0$ for all $i,j$, the exponential factorises term by term, $\exp[it\mathbf{H}]=\prod_i\exp[itT_i]$, and regrouping the local factors block by block gives, for every $t\in\mathbb{R}$,
\begin{equation}\label{eq:multi-sum}
\exp[it\mathbf{H}] =
e^{it\gamma}
\bigotimes_{k=1}^{n}\Big[\prod_{i=1}^{N}\exp[it\,\delta_i^{(k)}\,A_i^{(k)}]\Big],
\end{equation}
where $\gamma$ collects the all-scalar terms $T_i=\prod_l\lambda_i^{(l)}\,I$, $\delta_i^{(k)}=\prod_{l\ne k}\lambda_i^{(l)}$ if $A_i^{(k)}$ is the unique non-scalar factor of a nonzero $T_i$ (with $A_i^{(l)}=\lambda_i^{(l)}I$ for $l\ne k$), and $\delta_i^{(k)}=0$ otherwise; within a block, commuting summands have commuting non-scalar factors, so the order of the product is immaterial. No converse is claimed for an arbitrary commuting simple-tensor decomposition: cancellations can hide nonlocal summands even in the commuting case, $Z\otimes Z+(-Z\otimes Z)=0$ generates a separable flow although each displayed summand has two non-scalar factors. Theorem~\ref{thm:pauli-schmidt} gives the presentation-independent test.
\end{remark}

\subsection{Single-time resonances}\label{ssec:pauli}

\begin{remark}\label{rem:periodicity}
A single-time product unitary does not imply all-time product dynamics. With $A=\pi\sigma_Z$, $B=\sigma_Z$ on $\mathbb{C}^2\otimes\mathbb{C}^2$, one has
\[
\exp[iA\otimes B] = \mathrm{diag}(e^{i\pi},e^{-i\pi},e^{-i\pi},e^{i\pi}) = -I = (-I)\otimes I,
\]
which is separable although neither factor is scalar. The phases agree modulo $2\pi$ at this time. The following theorem characterises these resonances through the products of spectral gaps.
\end{remark}

\begin{theorem}[Single-time bipartite criterion]\label{thm:bipartite-single-t}
Let $A\in\cB_s(\cH_1)$, $B\in\cB_s(\cH_2)$ be self-adjoint on finite-dimensional spaces, and let $t\in\mathbb{R}$. Denote the distinct eigenvalues of $A$ by $\alpha_1,\dots,\alpha_p$ and those of $B$ by $\beta_1,\dots,\beta_q$. There exist unitaries $C\in\cU(\cH_1)$, $D\in\cU(\cH_2)$ satisfying $\exp[itA\otimes B]=C\otimes D$ (exact equality; any global phase can be absorbed into either factor, consistent with the up-to-phase convention of \S\ref{sec:prelims}) if and only if
\begin{equation}\label{eq:periodicity}
t(\alpha_i-\alpha_{i'})(\beta_j-\beta_{j'})\;\in\;2\pi\mathbb{Z}
\quad\text{for all }i,i'\in\{1,\dots,p\},\;j,j'\in\{1,\dots,q\}.
\end{equation}
In particular, \eqref{eq:periodicity} holds for every $t\in\mathbb{R}$ if and only if $A\in\mathbb{R}I$ or $B\in\mathbb{R}I$, recovering the all-time condition of Corollary~\ref{thm:bipartite}.
\end{theorem}

\noindent  The single-time factorisation question reduces to whether the bilinear phase table $\{t\alpha_i\beta_j\}_{ij}$ can be decomposed modulo $2\pi$ into a sum of row contributions $\phi_i$ and column contributions $\psi_j$.  For all $t$ outside a discrete resonant set this fails whenever both factors are non-scalar; it succeeds at special \emph{resonant} times when every pairwise interaction phase $t(\alpha_i-\alpha_{i'})(\beta_j-\beta_{j'})$ happens to fall on the $2\pi\mathbb{Z}$ lattice.

\begin{proof}
Let $A=\sum_i\alpha_i P_i$ and $B=\sum_j\beta_j Q_j$ be spectral decompositions with distinct eigenvalues and mutually orthogonal projections.

\smallskip\noindent\emph{Necessity.}  Since $A\otimes B$ commutes with both $A\otimes I$ and $I\otimes B$, the operator $\exp[itA\otimes B]$ does as well.  If $C\otimes D=\exp[itA\otimes B]$, then $(CA-AC)\otimes D=0$ and $C\otimes(DB-BD)=0$; since $C,D$ are invertible, $CA=AC$ and $DB=BD$.  Hence $C$ maps each eigenspace $P_i\cH_1$ to itself and $D$ maps each $Q_j\cH_2$ to itself; write $C_i:=C|_{P_i\cH_1}$ and $D_j:=D|_{Q_j\cH_2}$; as restrictions of the unitaries $C,D$ to invariant subspaces, $C_i$ and $D_j$ are themselves unitary.

Since $C$ preserves $P_i\cH_1$ and $D$ preserves $Q_j\cH_2$, the product $C\otimes D$ preserves the joint eigenspace $P_i\cH_1\otimes Q_j\cH_2$ and restricts there to $C_i\otimes D_j$. As $A\otimes B$ acts as the scalar $\alpha_i\beta_j$ on this subspace, the product constraint gives
\[
C_i\otimes D_j \;=\; e^{it\alpha_i\beta_j}\,I_{P_i\cH_1}\otimes I_{Q_j\cH_2}.
\]
We use the elementary fact that $M\otimes N=\lambda I$, with $\lambda\ne0$, forces both factors to be scalar. Choose a linear functional $f$ with $f(N)\ne0$ and apply the slice map $\mathrm{id}\otimes f$. Then $f(N)M=\lambda f(I)I$, so $M$ is a nonzero scalar multiple of the identity; substitution gives the same conclusion for $N$. Applied to the restrictions above, this gives $C_i=e^{i\phi_i}I_{P_i\cH_1}$ and $D_j=e^{i\psi_j}I_{Q_j\cH_2}$ for real phases $\phi_i,\psi_j$ with
\begin{equation}\label{eq:phi-psi-matching}
\phi_i+\psi_j\;\equiv\;t\alpha_i\beta_j\pmod{2\pi}\quad\text{for all }i,j.
\end{equation}
Adding the $(i,j)$ and $(i',j')$ instances of~\eqref{eq:phi-psi-matching} and subtracting the $(i,j')$ and $(i',j)$ instances, the $\phi$ and $\psi$ terms cancel on the left and the right-hand side collapses to $t(\alpha_i-\alpha_{i'})(\beta_j-\beta_{j'})$, which therefore lies in $2\pi\mathbb{Z}$; this is~\eqref{eq:periodicity}.

\smallskip\noindent\emph{Sufficiency.}  Fix reference indices $i_0$ and $j_0$.  For every $j$, choose a real representative $\psi_j$ satisfying $\psi_j\equiv t\alpha_{i_0}\beta_j\pmod{2\pi}$.  For every $i$, choose a real representative $\phi_i$ satisfying
\[
\phi_i\;\equiv\;t(\alpha_i-\alpha_{i_0})\beta_{j_0}\pmod{2\pi}.
\]
The value of $\phi_i$ is independent modulo $2\pi$ of the chosen reference $j_0$, because condition~\eqref{eq:periodicity} gives $t(\alpha_i-\alpha_{i_0})(\beta_{j_0}-\beta_{j_0'})\in2\pi\mathbb{Z}$ for all $j_0,j_0'$.  Then for every $i,j$:
\[
\phi_i+\psi_j-t\alpha_i\beta_j
\;\equiv\;t(\alpha_i-\alpha_{i_0})\beta_{j_0}+t\alpha_{i_0}\beta_j-t\alpha_i\beta_j
\;=\;-t(\alpha_i-\alpha_{i_0})(\beta_j-\beta_{j_0})\pmod{2\pi},
\]
and the right-hand side lies in $2\pi\mathbb{Z}$ by~\eqref{eq:periodicity}, so~\eqref{eq:phi-psi-matching} holds.  Define $C:=\sum_i e^{i\phi_i}P_i$ and $D:=\sum_j e^{i\psi_j}Q_j$; both are unitary and $(C\otimes D)|_{P_i\cH_1\otimes Q_j\cH_2}=e^{it\alpha_i\beta_j}I$, hence $C\otimes D=\exp[itA\otimes B]$.

\smallskip\noindent\emph{All-$t$ equivalence.}  If~\eqref{eq:periodicity} holds for every $t\in\mathbb{R}$, then for any $i,i',j,j'$ the quantity $c:=(\alpha_i-\alpha_{i'})(\beta_j-\beta_{j'})$ satisfies $tc\in2\pi\mathbb{Z}$ for all $t$, hence $c=0$; otherwise $\{tc:t\in\mathbb{R}\}=\mathbb{R}$ would be contained in the discrete set $2\pi\mathbb{Z}$.  Thus $(\alpha_i-\alpha_{i'})(\beta_j-\beta_{j'})=0$ for all $i,i',j,j'$, which holds iff $p=1$ ($A$ is scalar) or $q=1$ ($B$ is scalar).
\end{proof}

\section{Approximate separability}\label{sec:approx}

When $\Hns\ne0$, independent evolution under $\Hsep$ approximates the target dynamics. We first bound this error in operator and Hilbert--Schmidt norms, then construct a first-order correction and relate the residual to short-time entanglement diagnostics. Throughout, $\cS$ is fixed and $\Hsep,\Hns$ are the projection pair of~\eqref{eq:Hsep-closed}, except where another additive reference is explicitly allowed.

\begin{theorem}\label{thm:approx}
Let $\mathbf{H}\in\cB_s(\cH)$ on finite-dimensional $\cH$, with projection pair $\Hsep,\Hns$ as in~\eqref{eq:Hsep-closed}. Then for every $t\in\mathbb{R}$,
\begin{equation}\label{eq:approx-op}
\big\|\exp[it\mathbf{H}]-\exp[it\Hsep]\big\|_\op\;\le\;|t|\cdot\|\Hns\|_\op.
\end{equation}
Here $\exp[it\Hsep]=\bigotimes_k\exp[itH_k]$, with $\Hsep=\sum_k I^{\otimes\overline{S_k}}\otimes H_k$ as in~\eqref{eq:S-additive}, is $\cS$-separable, so the block-parallel evolution approximates $\exp[it\mathbf{H}]$ with an error growing at most linearly in the time $|t|$ and in $\|\Hns\|_\op$.
\end{theorem}

\noindent Thus $|t|\,\|\Hns\|_\op\le\varepsilon$ is a sufficient condition for block-parallel evolution to meet an operator-norm tolerance $\varepsilon$, assuming the block evolutions are exact. The bound has no additional factor depending on the module dimensions.

\begin{proof}
For self-adjoint $A,B$, Duhamel's identity, equivalently the standard integral representation for differences of matrix exponentials~\cite[Ch.~10]{Higham2008}, gives
\[ \exp[itA]-\exp[itB] = i\int_0^t \exp[i(t-s)A]\,(A-B)\,\exp[isB]\,ds. \]
With $A=\mathbf{H}$ and $B=\Hsep$, the integrand has operator norm at most $\|\Hns\|_\op$ because both exponential factors are unitary. The orientation reversal at negative $t$ is absorbed by the norm: $\big\|\int_0^t g(s)\,ds\big\|_\op\le|t|\sup_{s\in[\min(0,t),\max(0,t)]}\|g(s)\|_\op$ for every $t\in\mathbb{R}$, so taking norms gives~\eqref{eq:approx-op}.
\end{proof}

\begin{remark}[Choice of the additive reference]\label{rem:hs-vs-op}
The Hilbert--Schmidt projection $\Hsep$ is not, in general, the operator-norm-closest $\cS$-additive Hamiltonian, but it is within a fixed factor of it. By the closed form~\eqref{eq:Hsep-closed}, $\Pi_{\mathrm{ns}}=\mathrm{id}-\sum_k\mathcal E_k+(m-1)\tau$ is a sum of $2m$ operator-norm contractions (conditional expectations and the normalised trace are unital completely positive maps) and annihilates every $\cS$-additive operator. Hence for every $\cS$-additive $\mathbf{H}'$,
\[
\|\Hns\|_\op=\|\Pi_{\mathrm{ns}}(\mathbf{H}-\mathbf{H}')\|_\op\le 2m\,\|\mathbf{H}-\mathbf{H}'\|_\op ,
\]
and the bound~\eqref{eq:approx-op} is at most a factor $2m$ ($4$ for a bipartition) above what the best additive reference could give.
\end{remark}

\begin{proposition}[Hilbert--Schmidt refinement]\label{prop:HS}
Under the hypotheses of Theorem~\ref{thm:approx},
\begin{equation}\label{eq:approx-HS}
\big\|\exp[it\mathbf{H}]-\exp[it\Hsep]\big\|_\HS \le |t|\cdot\|\Hns\|_\HS.
\end{equation}
In the dimensionless form appropriate to unitaries ($\|U\|_\HS=\sqrt d$), and for qubit registers in terms of the cross-block Pauli coefficients,
\[ \frac{\|\exp[it\mathbf{H}]-\exp[it\Hsep]\|_\HS}{\sqrt d}\le |t|\,\frac{\|\Hns\|_\HS}{\sqrt d}=|t|\Big(\sum_{P:\,w_\cS(P)\ge2}c_P^2\Big)^{1/2}; \]
equivalently, for $\mathbf{H}\ne0$, $\|\exp[it\mathbf{H}]-\exp[it\Hsep]\|_\HS/\|\mathbf{H}\|_\HS\le|t|\sqrt{\rhoH(\mathbf{H})}$.
\end{proposition}

\begin{proof}
See Appendix~\ref{app:HS}.
\end{proof}

\subsection{Schmidt-rank refinement}\label{ssec:schmidt-refinement}

Theorem~\ref{thm:approx} bounds the worst-case operator-norm error using only the scalar quantity $\|\Hns\|_\op$. When $\Hns$ has additional bipartite structure, specifically low operator-Schmidt rank across $\cS$, the leading-order error inherits that structure.

For a bipartition $\cS=\{S_1,S_2\}$, the \emph{operator-Schmidt decomposition} of $M\in\cB(\cH_{S_1}\otimes\cH_{S_2})$ is
\begin{equation}\label{eq:osr-def}
M = \sum_{k=1}^{r} \sigma_k\, A_k\otimes B_k,
\end{equation}
with $\sigma_1\ge\dots\ge\sigma_r>0$, and $\{A_k\}\subset\cB(\cH_{S_1})$, $\{B_k\}\subset\cB(\cH_{S_2})$ Hilbert--Schmidt-orthonormal sets. The minimum such $r$ is the \emph{operator-Schmidt rank} $\mathrm{osr}_\cS(M)$, equivalently the matrix rank of the bipartite reshape $M_{(ai)(bj)}=\langle a,i|M|b,j\rangle\mapsto M_{(ab)(ij)}$, where $a,b$ index a basis of $\cH_{S_1}$ and $i,j$ a basis of $\cH_{S_2}$. We use the convention $\mathrm{osr}_\cS(0)=0$, with the empty Schmidt sum when $M=0$.

\begin{theorem}[Schmidt-rank refinement of Theorem~\ref{thm:approx}]\label{thm:schmidt-refinement}
Let $\mathbf{H}\in\cB_s(\cH_{S_1}\otimes\cH_{S_2})$ on a bipartite finite-dimensional Hilbert space admit a decomposition $\mathbf{H}=\Hsep+\Hns$ into self-adjoint parts, with $\Hsep$ $\cS$-additive and $\Hns$ admitting the operator-Schmidt decomposition~\eqref{eq:osr-def} of rank $r$. The canonical choice is~\eqref{eq:Hsep-closed}, but any such splitting is allowed in this theorem. Define
\begin{equation}\label{eq:Tt-def}
T_t \;:=\; it\,\Hns\,\exp[it\Hsep]\;=\; it\sum_{k=1}^{r}\sigma_k\,\big(A_k\,\exp[itH_1]\big)\otimes\big(B_k\,\exp[itH_2]\big),
\end{equation}
where $\Hsep=H_1\otimes I + I\otimes H_2$ with $H_1\in\cB_s(\cH_{S_1})$, $H_2\in\cB_s(\cH_{S_2})$. Then for every $t\in\mathbb{R}$:
\begin{enumerate}[\rm (i)]
\item \emph{Schmidt-rank propagation:} $\mathrm{osr}_\cS(T_t)=r$ for $t\ne0$ and $\mathrm{osr}_\cS(T_0)=0$.
\item \emph{First-order match:} $\|T_t\|_\op = |t|\,\|\Hns\|_\op$, recovering Theorem~\ref{thm:approx}'s bound (for the canonical splitting; the Duhamel argument applies verbatim to any $\cS$-additive reference generator).
\item \emph{Second-order residual:} the residual $R_t := \exp[it\mathbf{H}]-\exp[it\Hsep]-T_t$ satisfies the commutator-form estimate
\begin{equation}\label{eq:Rt-comm-bound}
\|R_t\|_\op \;\le\; \frac{t^2}{2}\,\big\|[\Hsep,\Hns]+\Hns^2\big\|_\op,
\end{equation}
and hence the coarser bound
\begin{equation}\label{eq:Rt-bound}
\|R_t\|_\op \;\le\; \frac{t^2}{2}\,\|\Hns\|_\op\,\big(\|\Hns\|_\op + 2\,\|\Hsep\|_\op\big).
\end{equation}
\end{enumerate}
\end{theorem}

\noindent The approximation $\exp[it\Hsep]+T_t$ therefore has a quadratic error bound, with exactly $r$ product terms in its correction for $t\ne0$. This is an algebraic approximation and need not be unitary; its implementation is discussed in Remark~\ref{rem:schmidt-knitting-cost}.

\noindent\emph{Proof strategy.} Duhamel's identity is integrated once by parts at the right endpoint, which extracts the term $T_t=it\,\Hns\exp[it\Hsep]$ and leaves a residual whose derivative contains only $[\Hsep,\Hns]$ and $\Hns^2$, giving~\eqref{eq:Rt-comm-bound} and then~\eqref{eq:Rt-bound}. Since $\exp[it\Hsep]=\exp[itH_1]\otimes\exp[itH_2]$, right multiplication by this local unitary preserves the operator-Schmidt rank, and the scalar $it$ only collapses the rank at $t=0$; unitary invariance of the operator norm gives $\|T_t\|_\op=|t|\|\Hns\|_\op$. The complete proof is given in Appendix~\ref{app:schmidt-refinement}.

\begin{remark}\label{rem:schmidt-corrections}
For a multipartite partition $\cS=\{S_1,\dots,S_m\}$ with $m\ge3$, the bounds~\eqref{eq:Rt-comm-bound}--\eqref{eq:Rt-bound} still hold: their proof uses only self-adjointness and $\mathbf{H}=\Hsep+\Hns$. For each bipartition separating one block from its complement, $T_t$ and $\Hns$ also have the same operator-Schmidt rank when $t\ne0$, since right multiplication by $\exp[it\Hsep]$ is local across that cut. These per-cut matrix ranks give lower bounds on the minimum number of multipartite product terms, but do not determine that number in general.
\end{remark}

\begin{remark}\label{rem:schmidt-knitting-cost}
The correction takes the form $\exp[it\Hsep]+T_t=(I+it\Hns)\exp[it\Hsep]$. A linear-combination-of-unitaries (LCU), quasi-probability, or postprocessing protocol that realises this approximation inherits the residual bound~\eqref{eq:Rt-comm-bound}. For $\Hns\ne0$, the quadratic bound is smaller than the linear bound when $0<|t|<2\|\Hns\|_\op/\kappa$, where $\kappa=\|[\Hsep,\Hns]+\Hns^2\|_\op$. Since a scalar shift of $\mathbf{H}$ changes $R_t$ only by a phase, the coarser bound~\eqref{eq:Rt-bound} may use $\min_{c\in\mathbb{R}}\|\Hsep-cI\|_\op$. The rank specifies the correction's product-term count; sampling overhead also depends on the coefficients and the local decompositions used by the protocol.
\end{remark}

\subsection{Connections to nearby diagnostics}\label{ssec:connections}

The same residual determines leading short-time entanglement diagnostics across a bipartition. We give the state-space and operator-space entangling-power coefficients, the Choi tripartite-information expansion, and the identification with the Zanardi--Dallas--Andreadakis (ZDA) generalised tensor-product-structure (g-TPS) framework~\cite{Zanardi2026}. Until the Choi-leg notation is introduced, write the blocks as $A,B$, with $d_A=\dim\cH_A$, $d_B=\dim\cH_B$, and $d=d_Ad_B$.\footnote{ADZ denotes the operator-space entangling-power work of Andreadakis--Dallas--Zanardi~\cite{Andreadakis2024}; ZDA denotes the g-TPS and algebraic out-of-time-ordered-correlator (A-OTOC) work of Zanardi--Dallas--Andreadakis~\cite{Zanardi2026}.}

\paragraph{State-space entangling power.} The entangling power of Zanardi, Zalka, and Faoro~\cite{Zanardi2000} is
\[ e_p(U) = \mathbb{E}\,\big[E_{\rm st}\big(U(\ket{\phi_1}\otimes\ket{\phi_2})\big)\big], \]
where the expectation is over uniformly random product states $\ket{\phi_1}\otimes\ket{\phi_2}$ (Haar-random on each local Hilbert space) and $E_{\rm st}(\ket\psi)=1-\tr(\varrho_A^2)$ is the linear entanglement entropy of a state (with $\varrho_A$ its reduced state on $\cH_A$). At short times it is set exactly by the non-separable mass, with a dimensional constant that differs from the operator-space one below.

\begin{proposition}[State-space entangling power, short time]\label{prop:ep-exact}
For finite-dimensional $\mathbf{H}\in\cB_s(\cH_A\otimes\cH_B)$,
\begin{equation}\label{eq:ep-exact}
e_p\big(\exp[it\mathbf{H}]\big) \;=\; \frac{2t^2}{(d_A+1)(d_B+1)}\,\|\Hns\|_\HS^2 \;+\; O(t^3).
\end{equation}
\end{proposition}

\begin{proof}
See Appendix~\ref{app:ep-exact}: local unitaries do not change $E_{\rm st}$, so $\exp[it\mathbf{H}]$ may be replaced by $\exp[-it\Hsep]\exp[it\mathbf{H}]=I+it\Hns+O(t^2)$; the linear entropy of a near-product state is $2t^2\|(Q_1\otimes Q_2)\Hns\ket{\phi_1\phi_2}\|^2+O(t^3)$ with $Q_i=I-\ket{\phi_i}\bra{\phi_i}$; and the Haar second-moment identity evaluates this quadratic form on the traceless$\,\otimes\,$traceless sector.
\end{proof}

\noindent Numerical checks of~\eqref{eq:ep-exact} are reported in \S\ref{ssec:auxiliary}.

\paragraph{Operator-space entangling power.} Write $\tr_A,\tr_B$ for the partial traces over $\cH_A,\cH_B$. For a unitary $U$ let $\mathcal{U}(\bullet)=U\bullet U^\dagger$ denote its adjoint channel (the calligraphic $\mathcal{U}$ is this superoperator, distinct from the unitary group $\cU(\cH)$ of \S\ref{sec:prelims}); for the flow we write $\mathcal{U}_t(\bullet)=\exp[it\mathbf{H}]\bullet\exp[-it\mathbf{H}]$. The central operational quantity is the \emph{operator-space} entangling power of ADZ~\cite{Andreadakis2024}: $E_p^{\op}(\mathcal{U}) := \mathbb{E}_{V_A,W_B}\,E_{\rm lin}\big(\mathcal{U}(V_A\otimes W_B)\big)$, where the expectation is over Haar-random local unitaries $V_A\in\cU(\cH_A)$, $W_B\in\cU(\cH_B)$, and $E_{\rm lin}$ is the linear operator entanglement of~\cite{Zanardi2001,Andreadakis2024}: $E_{\rm lin}(M)=1-\tr(\varrho_M^2)$, with $\varrho_M$ the reduced state, across the $(A,B)$ operator factorisation, of the normalised vectorisation $|M\rangle/\|M\|_\HS$. Because $E_{\rm lin}$ is a polynomial in the reduced operator-state purity, the short-time expansion below is analytic in $t$, with no non-analytic $t^2\ln(1/|t|)$ term (in contrast to the rank-deficient Choi reductions discussed later). The short-time formula of ADZ~\cite[Eq.~(14)]{Andreadakis2024} gives $E_p^{\op}(\mathcal{U}_t)=4(t/\tau_s)^2+O(t^3)$ with Gaussian scrambling rate $\tau_s^{-1}=d^{-1/2}\|\mathbf{H}_\perp\|_\HS$~\cite[Eq.~(12)]{Andreadakis2024}, a rate that originates in the minimal-scrambling programme of~\cite[Eq.~(14)]{Zanardi2024}; here $\mathbf{H}_\perp$ is the centred component of $\mathbf{H}$, written in global-phase-invariant form:
\begin{equation}\label{eq:tau-s-andreadakis}
\mathbf{H}_\perp \;=\; \mathbf{H} - \tfrac{I_A}{d_A}\!\otimes\!\tr_A(\mathbf{H}) - \tr_B(\mathbf{H})\!\otimes\!\tfrac{I_B}{d_B} + \tfrac{\tr(\mathbf{H})}{d}\,I.
\end{equation}
ADZ write~\eqref{eq:tau-s-andreadakis} without the last term; since the adjoint flow, and hence $E_p^{\op}$, is unchanged under $\mathbf{H}\mapsto\mathbf{H}+\alpha I$, their formula applies to the traceless representative, and~\eqref{eq:tau-s-andreadakis} is its shift-invariant form, equal to theirs when $\tr\mathbf{H}=0$. Lemma~\ref{lem:Hns-equals-Hperp} identifies $\mathbf{H}_\perp$ with the residual $\Hns$.

\begin{lemma}\label{lem:Hns-equals-Hperp}
For any finite-dimensional $\mathbf{H}\in\cB_s(\cH_A\otimes\cH_B)$, the centred component~\eqref{eq:tau-s-andreadakis} equals the residual $\Hns$ for $\cS=\{A,B\}$. Consequently,
\begin{equation}\label{eq:Hns-norm}
\|\Hns\|_\HS^2 \;=\; \rhoH(\mathbf{H})\,\|\mathbf{H}\|_\HS^2,
\end{equation}
including $\mathbf{H}=0$ under the convention $\rhoH(0)=0$.
\end{lemma}

\begin{proof}
For $m=2$, the right-hand side of~\eqref{eq:tau-s-andreadakis} is precisely $\mathbf{H}-\mathcal E_A(\mathbf{H})-\mathcal E_B(\mathbf{H})+\tau(\mathbf{H})=\Hns$ by~\eqref{eq:Hsep-closed}. Equation~\eqref{eq:Hns-norm} follows from the definition of $\rhoH$.
\end{proof}

\begin{proposition}[Operational identity for $\rhoH$]\label{prop:rho-Ep}
For any bipartite Hamiltonian $\mathbf{H}\in\cB_s(\cH_A\otimes\cH_B)$ with $d=d_Ad_B$, the operator-space entangling power of the flow $\mathcal{U}_t(\bullet)=\exp[it\mathbf{H}]\bullet\exp[-it\mathbf{H}]$ admits the short-time expansion
\begin{equation}\label{eq:rho-Ep}
E_p^{\op}\!\big(\mathcal{U}_t\big) \;=\; \frac{4t^2}{d}\,\|\Hns\|_\HS^2 \;+\; O(t^3) \;=\; \frac{4t^2}{d}\,\rhoH(\mathbf{H})\,\|\mathbf{H}\|_\HS^2 \;+\; O(t^3).
\end{equation}
Equivalently, $(d/4)$ times the leading $t^2$ coefficient of $E_p^{\op}(\mathcal{U}_t)$ is the partition-aware non-separable mass, with the ADZ Gaussian operator-scrambling rate $\tau_s^{-1}=\|\Hns\|_\HS/\sqrt{d}$.
\end{proposition}

\begin{proof}
Apply the ADZ expansion $E_p^{\op}(\mathcal{U}_t)=4(t/\tau_s)^2+O(t^3)$~\cite[Eq.~(14)]{Andreadakis2024} to the traceless representative of $\mathbf{H}$. Its Gaussian rate satisfies $\tau_s^{-2}=\|\mathbf{H}_\perp\|_\HS^2/d$~\cite[Eq.~(12)]{Andreadakis2024}. Substitution of Lemma~\ref{lem:Hns-equals-Hperp} gives both equalities in~\eqref{eq:rho-Ep}.
\end{proof}

\noindent
Section~\ref{ssec:exp6} tests the coefficient $4/d$ and the finite-time departure from this leading term.

\begin{remark}
Proposition~\ref{prop:rho-Ep} identifies the non-separable mass $\|\Hns\|_\HS^2=\rhoH\|\mathbf{H}\|_\HS^2$ with $(d/4)$ times the leading coefficient of the ADZ operator-space entangling power. For qubit registers, Andreadakis--Zanardi~\cite[Theorem~1]{Andreadakis2026} establish a complementary zero-set correspondence: $E_{\rm lin}(U^\dagger PU)=0$ for every Pauli string $P$ if and only if $U^\dagger=(V_A\otimes W_B)\,C$ with $C$ a Clifford circuit and $V_A,W_B$ local unitaries, so generating operator entanglement on Pauli strings is the exact operator-space signature of nonlocal nonstabilizerness generation. This is the same operator-entanglement geometry whose Haar-local average fixes $\|\Hns\|_\HS^2$, there averaged instead over Pauli strings.
\end{remark}

\paragraph{Tripartite-information identity.}
Choi-state scrambling gives a second diagnostic, based on the tripartite information of Hosur--Qi--Roberts--Yoshida~\cite{Hosur2016}, and at short times it too is governed exactly by $\Hns$. Let $\cS=\{S_1,S_2\}$ be a bipartition of $\{1,\dots,n\}$ and let $|U_t\rangle=d^{-1/2}\mathrm{vec}(U_t)\in\cH^{\otimes 2}$ denote the normalised Choi vector ($\mathrm{vec}$ stacks the columns of its argument, so the input legs come first) of $U_t=\exp[it\mathbf{H}]$. Aligning the input/output factorisations to $\cS$, write $A=S_1^{\mathrm{in}}$, $B=S_2^{\mathrm{in}}$, $C=S_1^{\mathrm{out}}$, $D=S_2^{\mathrm{out}}$ (from here to the end of this subsection, and in the Choi-information fits of \S\ref{ssec:auxiliary}, $A,B,C,D$ are these Choi legs, following the convention of~\cite{Hosur2016} and overriding the block shorthand of the preceding paragraphs). With the standard convention $I_3(A{:}C{:}D)=I(A{:}C)+I(A{:}D)-I(A{:}CD)$, where $I(X{:}Y)=S(\varrho_X)+S(\varrho_Y)-S(\varrho_{XY})$ is the quantum mutual information and $S(\varrho)=-\tr\varrho\ln\varrho$ the von Neumann entropy of the reduced Choi state (natural logarithm; all Choi-information quantities are in nats), scrambling makes $I_3$ negative, so we report the non-negative quantity $-I_3$. \begin{proposition}[Choi tripartite information, short time]\label{prop:choi-exact}
Let $\cS=\{S_1,S_2\}$ be a finite-dimensional bipartition, $\mathbf{H}\in\cB_s(\cH)$ with $\Hns\ne0$, and let $\sigma_1,\dots,\sigma_r$ be the operator-Schmidt coefficients of $\Hns$ across the cut, $p_k:=\sigma_k^2/\|\Hns\|_\HS^2$ and $H_{\mathrm{Sh}}(p):=-\sum_kp_k\ln p_k$. Set $\varepsilon_t:=t^2\|\Hns\|_\HS^2/d$. Then, as $t\to0^{+}$ (both sides are even in $t$),
\begin{equation}\label{eq:I3-leading}
 -I_3(A{:}C{:}D) \;=\; S(\varrho_{AC}) + O(t^4)
 \;=\; \varepsilon_t\Big[\,2\ln\frac1t + 1 + H_{\mathrm{Sh}}(p) + \ln\frac{d}{\|\Hns\|_\HS^2}\Big] + O\big(t^3\ln(1/t)\big),
\end{equation}
where $S(\varrho_{AC})$ is the von Neumann operator entanglement entropy of $U_t$ across the cut. Equivalently, for $t>0$ the normalised ratio $c(t):=-I_3(A{:}C{:}D)/[t^2\rhoH(\mathbf{H})\,d^{-1}\|\mathbf{H}\|_\HS^2]$, formed with the normalised mass $d^{-1}\|\mathbf{H}\|_\HS^2$ (equal to $\sum_Pc_P^2$ for qubit registers), satisfies $c(t)=2\ln(1/t)+a+o(1)$ with $a=1+H_{\mathrm{Sh}}(p)+\ln(d/\|\Hns\|_\HS^2)$.
\end{proposition}

\begin{proof}
See Appendix~\ref{app:choi-exact}. The input and output marginals of a unitary's Choi state are maximally mixed, which reduces $-I_3$ to $S(\varrho_{AC})+S(\varrho_{AD})-\ln d$; $\varrho_{AC}$ has eigenvalues $s_k(U_t)^2/d$ with $s_k$ the operator-Schmidt coefficients of $U_t$, which are $\sqrt d,\ t\sigma_1,\dots,t\sigma_r$ up to $O(t^2)$; and the first-order correction to $\varrho_{AD}$ vanishes because $\Hns$ is Hermitian.
\end{proof}

\noindent
The $t^2\ln(1/t)$ term comes from the $O(t^2)$ eigenvalues that emerge from the rank-one reduction $\varrho_{AC}(0)=|\Phi^+\rangle\langle\Phi^+|_{AC}$, where $|\Phi^+\rangle_{AC}$ is maximally entangled between the input and output copies of $S_1$. In the normalised ratio $c(t)$, the coefficient of $\ln(1/t)$ is universally $2$. The offset $a$ depends on the Schmidt spectrum and strength of $\Hns$ and shifts by $-2\ln\lambda$ under $\mathbf{H}\mapsto\lambda\mathbf{H}$ for $\lambda>0$. Section~\ref{ssec:auxiliary} tests both coefficients instance by instance.

\paragraph{Mereological identification.} In the ZDA framework, a g-TPS is represented by a choice of factor algebra and its commutant. For the canonical full-factor bipartite g-TPS associated with the fixed hardware partition $\cS=\{S_1,S_2\}$, after choosing the tensor-product identification, the Hilbert--Schmidt projection off the factor algebra plus its commutant is exactly our non-separable projection.

\begin{proposition}[ZDA mereology identification]\label{prop:zda-identification}
Let $\cS=\{S_1,S_2\}$ be a finite-dimensional bipartition with the canonical full matrix factor algebra of this chosen tensor-product structure, and use the unnormalised Hilbert--Schmidt inner product $\langle M,N\rangle_\HS=\tr(M^\dagger N)$. Let
\[
A_{\rm fac}:=\cB(\cH_{S_1})\otimes I,\qquad
A_{\rm fac}' := I\otimes\cB(\cH_{S_2}),\qquad
L_0:=A_{\rm fac}+A_{\rm fac}'
\]
where $A_{\rm fac}'$ is the commutant factor. For $d_{S_1},d_{S_2}\ge2$ the linear space $L_0$ is not an algebra (it is not closed under multiplication); it is the complex span of the block-local operators obtained from the fixed g-TPS, with the shared identity direction counted once. It is adjoint-closed, its self-adjoint part is the $\cS$-additive subspace of \S\ref{sec:prelims}, and the Hilbert--Schmidt projection onto $L_0$ commutes with the adjoint and restricts on $\cB_s(\cH)$ to the projection defining $\Hsep$. Then:
\begin{enumerate}[\rm (i)]
\item Let $\Pi_{L_0}$ be the Hilbert--Schmidt orthogonal projection of $\cB(\cH)$ onto $L_0$, and set $Q_0:=\mathrm{id}-\Pi_{L_0}$ (the linear map on $\cB(\cH)$ projecting off the factor algebra plus commutant, as in ZDA). Then $Q_0$ restricted to $\cB_s(\cH)$ coincides with our projection $\Pi_{\mathrm{ns}}$ onto the orthogonal complement of the $\cS$-additive subspace (equivalently, $Q_0$ is the complex-linear extension of $\Pi_{\mathrm{ns}}$), so $Q_0(\mathbf{H})=\Hns$ for every $\mathbf{H}\in\cB_s(\cH)$.
\item The ZDA scrambling functional $\sigma_s(L_0,\mathbf{H}):=\|Q_0(\mathbf{H})\|_\HS$~\cite[Eq.~(4)]{Zanardi2026}, the Gaussian rate of their A-OTOC, which in the unnormalised convention is the Hilbert--Schmidt norm of the off-algebra projection, has square (we call $\sigma_s^2$ the scrambling variance)
\begin{equation}\label{eq:zda-sigma}
\sigma_s^2(L_0,\mathbf{H}) \;=\; \|Q_0(\mathbf{H})\|_\HS^2 \;=\; \|\Hns\|_\HS^2 \;=\; \rhoH(\mathbf{H})\,\|\mathbf{H}\|_\HS^2
\end{equation}
for $\mathbf{H}\ne0$, with the zero case again covered by $\rhoH(0)=0$. If the normalised Hilbert--Schmidt inner product $d^{-1}\tr(M^\dagger N)$ is used instead, both $\sigma_s^2$ and the squared norms in this line are divided by $d$, and the associated A-OTOC coefficient is rescaled consistently by that convention.
\item The ZDA A-OTOC associated with the factor algebra $A_{\rm fac}$ (equivalently with the subspace $L_0=A_{\rm fac}+A_{\rm fac}'$, which is all its short-time expansion uses), written $G_{L_0}(t)$~\cite[Eq.~(3)]{Zanardi2026}, has the short-time expansion $G_{L_0}(t)=(2/d)\sigma_s^2 t^2 + O(t^3)$, first derived in~\cite{Zanardi2024} (the proposition preceding their Eq.~(14), stated for $\tilde{\mathbf{H}}=\mathbf{H}/\sqrt d$) and restated with the explicit $2/d$ in~\cite{Zanardi2026}, where $\sigma_s^2$ is the unnormalised variance of part (ii) in ZDA's convention; together with Proposition~\ref{prop:rho-Ep} this gives
\[
E_p^{\op}(\mathcal{U}_t)=2G_{L_0}(t)+O(t^3),
\]
so the factor of $2$ is the ratio between the leading $t^2$ coefficients in ADZ's $E_p^{\op}$ ($4/d$) and ZDA's $G_{L_0}$ ($2/d$) definitions.
\end{enumerate}
\end{proposition}

\begin{proof}
Part (i) is the basis computation of Appendix~\ref{app:zda-identification}, using Lemma~\ref{lem:Hns-equals-Hperp}; part (ii) and a direct Haar-average derivation of the short-time coefficient in part (iii) are also given there.
\end{proof}

\noindent
Proposition~\ref{prop:zda-identification} gives the explicit Hamiltonian-level evaluation of the ZDA fixed-g-TPS scrambling projection for a hardware-dictated partition. ZDA vary the g-TPS at fixed $\mathbf{H}$, whereas here $\cS$ is fixed and $\mathbf{H}$ varies; both use the same Hilbert--Schmidt projection, whose common zero set is $\rhoH=0$.

\section{Numerical experiments}\label{sec:experiments}

We test the Pauli certificate and single-time resonances, the short-time scrambling coefficient, and the approximation and rank-refinement bounds, then benchmark block-factorised simulation. Further comparisons examine Choi information, finite-time entangling power, long-time operator growth, partition-aligned Trotterisation, and published cutting overheads. The exact-boundary examples serve as implementation checks; the remaining experiments test the quantitative bounds and their range of usefulness.

Throughout this section, $n$ is the number of qubits and $d=2^n$ is the Hilbert-space dimension; for one-dimensional chain benchmarks, $L$ or $N$ denotes the number of sites, following the notation of the work each benchmark reproduces ($L$ for the ADZ chain regimes, $N$ for the XYZ integrability chain). Timings are classical-simulator measurements on one node, using a central processing unit (CPU) and a graphics processing unit (GPU). Appendix~\ref{app:repro} gives the reproduction archive and timing setup.

\subsection{Check 1: exact-boundary detection}\label{ssec:exp-boundary}

\begin{figure}[!htbp]\centering
\figIfExists[width=74mm]{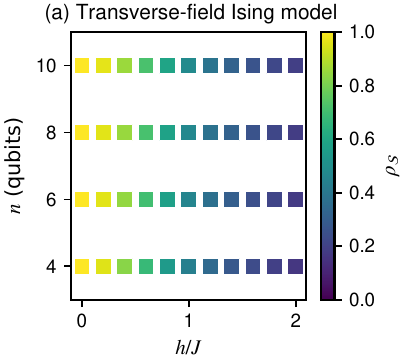}\hfill
\figIfExists[width=74mm]{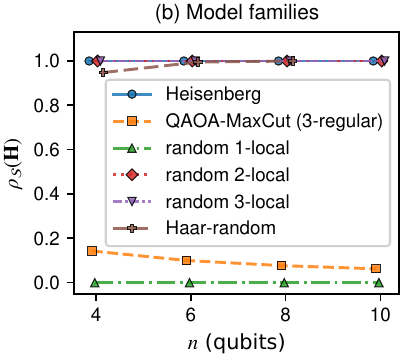}
\caption{Pauli--Schmidt landscape on the singleton partition. (a) Transverse-field Ising model landscape (discrete markers at integer $n\in\{4,6,8,10\}$ versus $h/J$, coloured by $\rhoH$); the non-separable mass fraction $\rhoH$ is carried entirely by the interaction term, decreases as the separable transverse-field contribution grows, and is nearly independent of $n$: on the open chain $\rhoH=(n-1)J^2/[(n-1)J^2+nh^2]$ varies by under $0.05$ across $n$ at fixed $h/J$ (it is exactly $1$ at $h/J=0$, and with periodic boundaries it would be the $n$-independent $J^2/(J^2+h^2)$). (b) Companion model families (markers horizontally offset for legibility; the Haar-random series stops at $n=8$). The $\rhoH=0$ series are exactly the all-$t$ separable instances of Theorem~\ref{thm:pauli-schmidt}.}\label{fig:exp1}
\end{figure}

\paragraph{Model-class detection.} Figure~\ref{fig:exp1} tabulates $\rhoH$ on the singleton partition for common qubit Hamiltonian families. The singleton partition treats each qubit as its own module, so a Pauli string contributes to $\Hns$ exactly when it acts nontrivially on at least two qubits. The displayed behaviour is a direct consequence of the Pauli support rule in Theorem~\ref{thm:pauli-schmidt}:
\begin{itemize}
\item Transverse-field Ising model (TFIM): $\mathbf{H}=-J\sum_{i=1}^{n-1} Z_iZ_{i+1}-h\sum_{i=1}^{n} X_i$ (open chain) has $\rhoH=1$ at the pure interaction point $h/J=0$, decreasing as the separable transverse-field term grows.
\item Isotropic Heisenberg chain (open): every term is two-site, so $\rhoH=1$ exactly.
\item Random 1-local Hamiltonians: $\rhoH=0$ exactly.
\item Random 2-local and 3-local Hamiltonians: sums over all Pauli strings acting nontrivially on exactly $k$ sites (block-weight $k$ on the singleton partition) with independent normal coefficients; every string is multi-qubit, so $\rhoH=1$ for every nonzero draw.
\item QAOA-MaxCut: the MaxCut cost operator used in the quantum approximate optimisation algorithm, $\mathbf{H}=\sum_{(i,j)\in E}\tfrac12(1-Z_iZ_j)$, on random $3$-regular graphs $G=(V,E)$. On the singleton partition every edge term is cross-block and the identity offset is the only $\cS$-additive contribution, so $\rhoH=1/(|E|+1)$ (the identity-offset dilution of Remark~\ref{rem:shift}; the traceless part gives $\rhoH=1$).
\item Unitarily invariant random Hermitian generators: concentrate near $\rhoH=1$ as dimension grows.
\end{itemize}

\paragraph{Periodicity counterexample and single-time resonance.} Corollary~\ref{thm:bipartite} requires the all-$t$ quantifier because a non-additive generator can produce separable unitary group elements at isolated resonant times (Remark~\ref{rem:periodicity}). For $\mathbf{H}=\pi\,\sigma_Z\otimes\sigma_Z$ with $\rhoH(\mathbf{H})=1$ on the standard bipartition,
\[
U_t=\exp[it\mathbf{H}]=\mathrm{diag}(e^{it\pi},e^{-it\pi},e^{-it\pi},e^{it\pi}).
\]
The single-time criterion of Theorem~\ref{thm:bipartite-single-t} predicts separability exactly when $4\pi t\in 2\pi\mathbb{Z}$, i.e.\ $t\in\tfrac{1}{2}\mathbb{Z}$. Thus $U_t=\pm I$ at integer times, and at $t=0.5$ one has $U_{0.5}=\sigma_Z\otimes(i\sigma_Z)$. At a non-resonant time such as $t=0.3$, the operator-Schmidt rank is $2$, so $U_{0.3}$ is not separable. On the grid $t\in\{0,0.1,\dots,1\}$, the operator-Schmidt rank of $U_t=\cos(\pi t)\,I+i\sin(\pi t)\,\sigma_Z\otimes\sigma_Z$ is $1$ exactly at $t\in\{0,0.5,1\}$ and $2$ elsewhere. This illustrates the distinction between all-time product dynamics and isolated separable group elements.

\subsection{Check 2: short-time operator-scrambling coefficient}\label{ssec:exp6}

\begin{figure}[!htbp]\centering
\figIfExists[width=95mm]{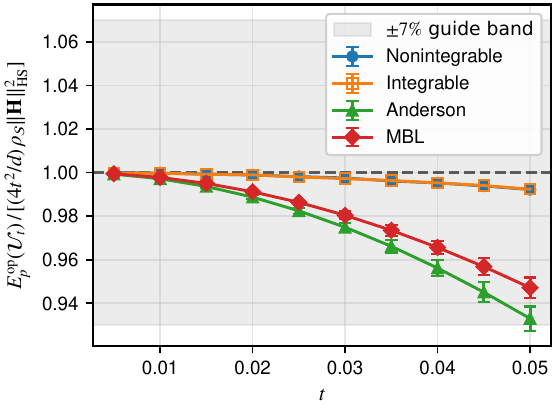}
\caption{Short-time check of the $4t^2/d$ coefficient of Proposition~\ref{prop:rho-Ep} at $L=8$. Per-realisation ratio $E_p^{\op}(\mathcal{U}_t)/[(4t^2/d)\,\rhoH\|\mathbf{H}\|_\HS^2]$ versus $t$; for the disordered Anderson and many-body-localised (MBL) regimes the markers are means over $32$ disorder realisations and the error bars are the standard error. The realisation-averaged ratio approaches $1$ as $t\to0$ for all four regimes and stays inside the shaded $\pm7\%$ guide band across $0<t\le0.05$. At $t=0.05$ the deviation is $\approx6.7\%\pm0.6\%$ (Anderson) and $\approx5.3\%\pm0.5\%$ (MBL), and below $1\%$ for the two clean Ising regimes, whose curves coincide at this scale; the residual is higher-order curvature (the relative deviation grows as $t^2$, consistent with $E_p^{\op}(\mathcal{U}_t)$ being even in $t$ for these real-symmetric chains).}\label{fig:exp6-short}
\end{figure}

Proposition~\ref{prop:rho-Ep} is a local-in-time statement:
\[
E_p^{\op}(\mathcal{U}_t)=(4t^2/d)\,\rhoH(\mathbf{H})\,\|\mathbf{H}\|_\HS^2+O(t^3).
\]
Check~1 located the boundary $\rhoH=0$; moving off it, we test the coefficient predicted by Proposition~\ref{prop:rho-Ep} on the four one-dimensional chain regimes of ADZ~\cite{Andreadakis2024} at $L=8$, across the balanced bipartition (sites $1$--$4\,|\,5$--$8$, so $d=2^{8}$); here $\rhoH$ is the cross-block non-separable mass fraction of the bipartite Proposition~\ref{prop:rho-Ep}.\footnote{All four use the chain $\mathbf{H}=-\sum_i Z_iZ_{i+1}-\sum_i(h_z\,Z_i+g_i\,X_i)$ (open boundary; here $h_z$ is a longitudinal field, unlike the transverse $h$ of the TFIM in Check~1): nonintegrable ($h_z=0.5$, $g_i=1.05$), integrable ($h_z=0$, $g_i=1$), Anderson ($h_z=0$, $g_i\sim\mathrm{Uniform}[-10,10]$), and many-body-localised (MBL; $h_z=0.5$, $g_i\sim\mathrm{Uniform}[-10,10]$). Here $E_p^{\op}=\mathbb{E}\,[1-\sum_k\lambda_k^2]$, with $\{\lambda_k\}$ the eigenvalues of the reduced operator state (the squared, normalised operator-Schmidt coefficients, $\sum_k\lambda_k=1$); $E_p^{\op}$, a local-Haar average, is bounded by $(d_{S_1}^2-1)(d_{S_2}^2-1)/(d^2-1)=255/257\approx0.992$ on this cut~\cite[Eq.~(15)]{Andreadakis2024}, so the saturation value $\approx0.99$ is near-maximal.} Figure~\ref{fig:exp6-short} plots the ratio $E_p^{\op}(\mathcal{U}_t)/[(4t^2/d)\rhoH(\mathbf{H})\|\mathbf{H}\|_\HS^2]$ (for the disordered regimes the ratio is formed per disorder realisation and then averaged, which avoids a product-of-means bias between $\rhoH$ and $\|\mathbf{H}\|_\HS^2$), which the identity predicts to approach $1$ as $t\to0$. All four regimes approach unity in the sampled short-time window; the largest deviations occur in the disordered regimes and are consistent with higher-order curvature beyond the leading $t^2$ coefficient.

\begin{figure}[!htbp]\centering
\figIfExists[width=95mm]{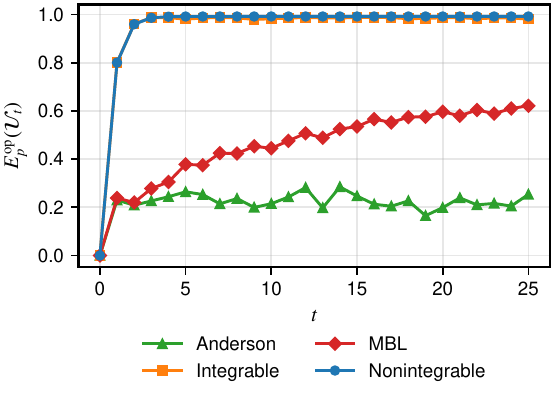}
\caption{Long-time operator-space entangling power of open Ising chains with $L=8$ sites across the balanced cut. Clean integrable and nonintegrable chains approach near-maximal values, whereas Anderson-localised and many-body-localised chains saturate lower. Times are sampled at integer steps over $0\le t\le25$; each disordered regime uses $16$ realisations, each clean regime one instance, and each evaluation $300$ local-Haar samples.}\label{fig:exp6-long}
\end{figure}

Figure~\ref{fig:exp6-long} repeats the measurement over $0\le t\le25$ to mark the limit of the theorem. Long-time saturation separates ordered from disordered regimes: both clean Ising regimes saturate near $E_p^{\op}\approx0.99$, while the two disordered regimes stay well below it, the non-interacting Anderson regime lowest (late-time mean $0.22$ over $t\in[15,25]$, averaged over $16$ disorder realisations) and the interacting MBL regime in between ($0.58$). The disordered means fluctuate strongly from realisation to realisation at this size, so the $16$-realisation average is what separates them; the quantitative check of the $4/d$ coefficient in Figure~\ref{fig:exp6-short} is carried to $32$ realisations.

The non-separable mass $\|\Hns\|_\HS^2=\rhoH(\mathbf{H})\|\mathbf{H}\|_\HS^2$ thus fixes the leading scrambling coefficient $(4/d)\|\Hns\|_\HS^2$, but not the later-time saturation value.

\subsection{Check 3: approximation-bound scaling, tightness, and Schmidt-rank propagation}\label{ssec:exp2}

On the balanced bipartition $\{1,\dots,n/2\}\,|\,\{n/2+1,\dots,n\}$, we test the linear perturbative error, the finite-time tightness of Theorem~\ref{thm:approx}, and the rank and residual of the first-order correction in Theorem~\ref{thm:schmidt-refinement}.

\begin{figure}[!htbp]\centering
\figIfExists[width=150mm]{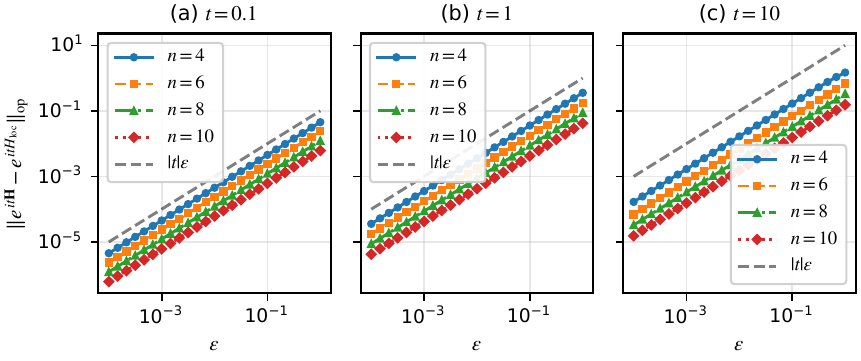}
\caption{Approximate-separability error $\|\exp[it\mathbf{H}]-\exp[itH_{\rm loc}]\|_\op$ vs.\ perturbation strength $\varepsilon$ at $n\in\{4,6,8,10\}$ and three values of $t$: (a) $0.1$, (b) $1$, (c) $10$. Here $\mathbf H=H_{\rm loc}+\varepsilon V$, $H_{\rm loc}$ is a sum of single-qubit terms, and $\|V\|_\HS=1$; markers are means over $10$ independent instances. Dashed: unit-slope guide proportional to $|t|\varepsilon$. Because $V$ is Hilbert--Schmidt-normalised, the Duhamel bound for this comparison is $|t|\,\varepsilon\|V\|_\op$; the dashed guide indicates the exponent rather than the exact operator-norm prefactor.}\label{fig:exp2}
\end{figure}

\paragraph{Scaling on random perturbations.} We construct $\mathbf{H}=H_{\rm loc}+\varepsilon V$, where $H_{\rm loc}$ is a random 1-local sum and $V$ is a self-adjoint perturbation from the Gaussian unitary ensemble (GUE) normalised to $\|V\|_\HS=1$ and held fixed across $\varepsilon$ and $t$ within an instance, and compare against the block evolution generated by $H_{\rm loc}$. The Duhamel argument behind Theorem~\ref{thm:approx} applies to any $\cS$-additive reference generator and bounds the plotted error by $|t|\,\varepsilon\|V\|_\op$: first order in $\varepsilon$ at fixed $V$. (Relative to the canonical projection $\Hsep=H_{\rm loc}+\varepsilon\,(V-\Pi_{\mathrm{ns}}V)$ the residual is $\Hns=\varepsilon\,\Pi_{\mathrm{ns}}V$, the cross-block part of the perturbation; both comparisons scale identically in $\varepsilon$.) Figure~\ref{fig:exp2} shows unit slope on log-log axes in the perturbative regime for $n\in\{4,6,8,10\}$ and $t\in\{0.1,1,10\}$; increasing $t$ raises the prefactor, while the small-$\varepsilon$ exponent remains one.

\begin{figure}[!htbp]\centering
\figIfExists[width=150mm]{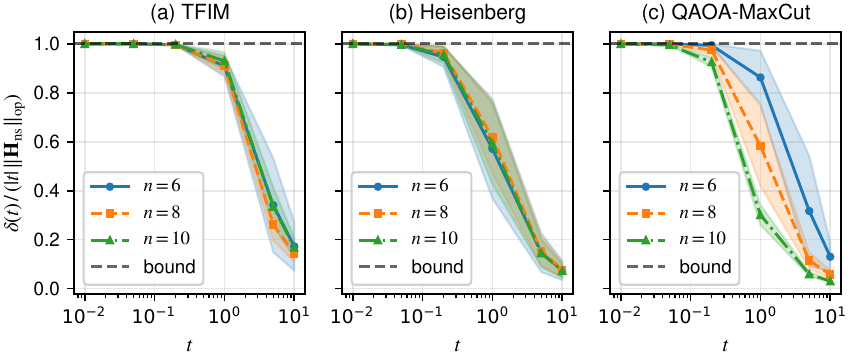}
\caption{Numerical tightness check for Theorem~\ref{thm:approx} on (a) transverse-field Ising chains, (b) Heisenberg chains and (c) MaxCut cost Hamiltonians used in the quantum approximate optimisation algorithm. The ratio $\delta(t)/(|t|\,\|\Hns\|_\op)$ approaches $1$ as $t$ decreases along the grid $t\in\{0.01,0.05,0.2,1,5,10\}$, for $n\in\{6,8,10\}$. Shaded bands show $\pm1$ population standard deviation over $5$ random instances. Finite-time cancellations make the error smaller than the linear norm bound.}\label{fig:exp9}
\end{figure}

\paragraph{Finite-time tightness on structured Hamiltonians.} For any fixed nonzero $\Hns$, Taylor expansion gives $\exp[it\mathbf{H}]-\exp[it\Hsep]=it\Hns+O(t^2)$, so the ratio below approaches $1$ as $t\to0$. We use three structured model families to confirm this analytically guaranteed limit at the implementation level and to show how the finite-time ratio departs from it: the TFIM above, the isotropic Heisenberg chain $\mathbf{H}=J\sum_{i=1}^{n-1}(X_iX_{i+1}+Y_iY_{i+1}+Z_iZ_{i+1})$ (open chain), and QAOA-MaxCut on an Erd\H{o}s--R\'enyi graph $G(n,1/2)$. For each instance we form
\[
\frac{\delta(t)}{|t|\,\|\Hns\|_\op},\qquad
\delta(t)=\|\exp[it\mathbf{H}]-\exp[it\Hsep]\|_\op .
\]
The theorem requires this ratio to be at most $1$. Table~\ref{tab:exp9} shows ranges over $n\in\{6,8,10\}$ and $5$ random instances per model; Figure~\ref{fig:exp9} plots the same ratios on the extended grid $t\in\{0.01,0.05,0.2,1,5,10\}$.

\begin{table}[!htbp]\centering\small
\begin{tabular}{lcccc}
\toprule
& $t=0.01$ & $t=0.05$ & $t=0.2$ & $t=1.0$ \\
\midrule
TFIM        & $1.0000$ & $0.9996$--$1.0000$ & $0.9943$--$0.9995$ & $0.8634$--$0.9879$ \\
Heisenberg  & $0.9997$--$1.0000$ & $0.9921$--$0.9995$ & $0.8888$--$0.9916$ & $0.3374$--$0.8312$ \\
QAOA-MaxCut & $0.9997$--$1.0000$ & $0.9933$--$1.0000$ & $0.8967$--$0.9996$ & $0.2494$--$0.9896$ \\
\bottomrule
\end{tabular}
\caption{Dimensionless tightness ratio $\delta(t)/(|t|\,\|\Hns\|_\op)$ for Theorem~\ref{thm:approx}. Within each model row, ranges aggregate over $n\in\{6,8,10\}$ and $5$ random instances (open-chain TFIM with $J\sim\mathrm{Uniform}[0.5,2]$, $h\sim\mathrm{Uniform}[0,1.5]$; open Heisenberg chain with $J\sim\mathrm{Uniform}[0.5,2]$; QAOA-MaxCut on an Erd\H{o}s--R\'enyi $G(n,1/2)$ graph). The ratio is bounded by $1$ as required and approaches $1$ as $t\to0$; the limit itself is universal (it follows from the Taylor expansion for every $\Hns\ne0$), so the table is an implementation-level check, and the informative content is the finite-$t$ departure.}\label{tab:exp9}
\end{table}

A second backend (GPU eigendecomposition, CuPy) on the same instances and grid reproduces every ratio in the table to within $4\times10^{-13}$.

\paragraph{Rank-propagation check.} Theorem~\ref{thm:schmidt-refinement} also identifies the product-term structure of the first-order correction added back to the block-parallel evolution. The correction is
\[
T_t=it\,\Hns\exp[it\Hsep]
\]
so right multiplication of $\Hns$ by the local unitary $\exp[it\Hsep]$ preserves operator-Schmidt rank while the remainder drops to second order. We generate sampled bipartite Hamiltonians $\mathbf{H}=\Hsep+\Hns$ with prescribed operator-Schmidt rank and evaluate
\[
R_t=\exp[it\mathbf{H}]-\exp[it\Hsep]-T_t.
\]
The floating-point computation reproduces the exact identities $\mathrm{osr}_\cS(T_t)=\mathrm{osr}_\cS(\Hns)$ (for $t\ne0$) and $\|T_t\|_\op=|t|\|\Hns\|_\op$ of Theorem~\ref{thm:schmidt-refinement}(i)-(ii) across the sampled ranks, sizes, and times; the empirical content is the second-order residual, which satisfies
\[
\|R_t\|_\op \le \tfrac{t^2}{2}\,\|\Hns\|_\op(\|\Hns\|_\op+2\|\Hsep\|_\op)
\]
and follows $\|R_t\|_\op\propto t^2$ to within $1\%$ at small $t$; the grid and instance counts are recorded in the data archive (Appendix~\ref{app:repro}). At $t\le0.01$, the ratio $\|R_t\|_\op/[(t^2/2)\kappa]$ for the commutator-form bound~\eqref{eq:Rt-comm-bound} rounds to $1.000$ on every instance, consistent with its asymptotic tightness. The coarse bound~\eqref{eq:Rt-bound} is looser by factors of $2.2$--$3.8$ (median $3.0$) on these instances.

\subsection{Check 4: block-factorised statevector simulation}\label{ssec:exp3}

When $\rhoH=0$, the factorisation $U_t=\bigotimes_k\exp[itH_k]$ allows a product input to be evolved as separate block vectors, using $\sum_k2^{|S_k|}$ amplitudes. Our timing benchmark instead materialises the full output vector for comparison with full-system baselines: it exponentiates each block Hamiltonian and applies the resulting unitaries by tensor contraction. Thus the measured timings exploit the factorised unitary while retaining full-statevector storage.

\begin{figure}[!htbp]\centering
\figIfExists[width=150mm]{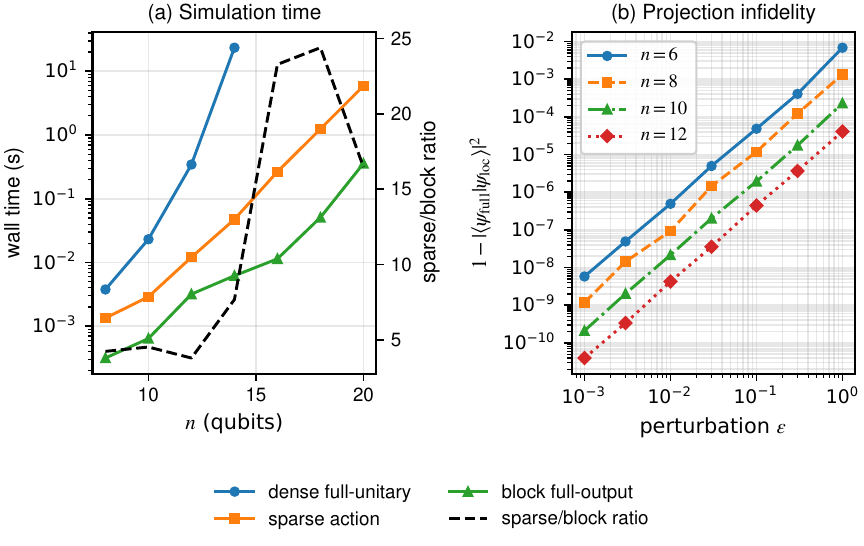}
\caption{Block-factorised product-state simulation. (a) Absolute wall-times (medians of five repetitions) vs.\ system size $n$ on a single Grace Hopper node; dense full-unitary evolution is shown only until the memory cap, while the sparse path and the block full-output path continue to $n=20$. The dashed curve gives the sparse/block timing ratio tabulated in Table~\ref{tab:exp3}. (b) Infidelity after replacing $\mathbf{H}=H_{\rm loc}+\varepsilon V$ by its unperturbed block-additive part $H_{\rm loc}$ (not the canonical $\Hsep$, which retains the block-local part of $\varepsilon V$), plotted against $\varepsilon$ for $n\in\{6,8,10,12\}$.}\label{fig:exp3}
\end{figure}

Each of the two balanced blocks is an open transverse-field Ising chain on $n/2$ qubits, with $J\sim\mathrm{Uniform}[0.5,2]$ and $h\sim\mathrm{Uniform}[0,1.5]$ at a fixed seed. No bond crosses the cut, so the evolution from $\ket{0}^{\otimes n}$ can entangle qubits within each block while preserving $\rhoH=0$. Table~\ref{tab:exp3} compares the block path with two baselines on one Grace Hopper node. The dense baseline forms the full unitary and requires $O(d^2)$ storage; the sparse baseline applies $\exp[it\mathbf{H}]$ directly to the full statevector using SciPy's \texttt{expm\_multiply}~\cite{AlMohy2011}. The latter permits comparison up to $n=20$ without forming the global unitary. Figure~\ref{fig:exp3} plots these timings and the error from an approximate block-additive projection.

\begin{table}[!htbp]\centering\small
\begin{tabular}{ccccc}
\toprule
$n$ & $d=2^n$ & dense time (s) & sparse/block ratio & block fidelity \\
\midrule
$8$  & $256$       & $0.004$ & $4.2\times$  & $1.000$ \\
$10$ & $1024$      & $0.023$ & $4.5\times$  & $1.000$ \\
$12$ & $4096$      & $0.345$ & $3.8\times$  & $1.000$ \\
$14$ & $16384$     & $23.47$ & $7.7\times$  & $1.000$ \\
$16$ & $65536$     & --     & $23.3\times$ & $1.000$ \\
$18$ & $262144$    & --     & $24.4\times$ & $1.000$ \\
$20$ & $1048576$   & --     & $16.5\times$ & $1.000$ \\
\bottomrule
\end{tabular}
\caption{Block-factorised product-state simulation versus monolithic statevector baselines on the balanced bipartition at $t=1.0$ (backend and memory details in Appendix~\ref{app:repro}). The dense path explicitly forms the full unitary and is skipped (dashes) once the conservative dense-memory estimate exceeds the preset cap. The sparse/block ratio compares full-system sparse exponential-action evolution (CPU) against the block path (block eigendecompositions on the GPU for block dimension $\ge64$, i.e.\ $n\ge12$, and on the CPU below that; tensor contraction on the CPU throughout), which reconstructs the full output vector for comparison; the CPU-only algorithmic comparison is reported in \S\ref{ssec:exp3}. Block fidelity is $|\langle\psi_{\rm sparse}|\psi_{\rm block}\rangle|^2$ on the same Hamiltonian instance. The instance has an open transverse-field Ising chain on each block and no cross bond (\S\ref{ssec:exp3}). Wall times are medians over five timing repetitions of the same Hamiltonian instance and ratios are ratios of those medians. At $n\le12$ the sparse and block paths each have median wall times within ${\sim}12$~ms, and at $n=10$ the five block-path repetitions are scattered over $0.6$--$19$~ms (a host-side timing scatter we did not root-cause), so the small-$n$ ratios are indicative only and the asymptotic comparison is carried by the $n\ge14$ rows.}\label{tab:exp3}
\end{table}

The block path forms only $m=2$ matrices of size $2^{n/m}\times2^{n/m}$ and reaches $n=20$, where the dense full-unitary baseline is no longer feasible under the memory cap. Because the Hamiltonian here is exactly $\cS$-additive ($\rhoH=0$), $U_t=\bigotimes_k\exp[itH_k]$ holds as an identity, so the block fidelity of $1.000$ to the sparse path checks backend agreement rather than the separability criterion. Same-device CPU-only medians in the data archive isolate an algorithmic speed-up of $3.3\times$ and $3.5\times$ at $n=18$ and $n=20$ (and $3.4\times$ at $n=16$, where the CPU-block time sits on a timing floor); the larger ratios in Table~\ref{tab:exp3} additionally include GPU acceleration of the block eigendecompositions and should not be read as a pure algorithmic gain.

The right panel of Figure~\ref{fig:exp3} shows the complementary approximate case: $\mathbf{H}=H_{\rm loc}+\varepsilon V$ on the same balanced bipartition, where $H_{\rm loc}=\sum_k I^{\otimes\overline{S_k}}\otimes H_k$ with each $H_k$ a dense random Hermitian matrix on its block (a different instance from the interacting TFIM blocks of Table~\ref{tab:exp3}) and $V$ is a random Hermitian perturbation containing both cross-block and block-local Pauli terms, normalised in Hilbert--Schmidt norm and drawn independently at each $\varepsilon$; input state $\ket{0}^{\otimes n}$, $t=1$, and error $1-|\langle\psi_{\rm full}|\psi_{\rm loc}\rangle|^2$, where $\ket{\psi_{\rm full}}:=\exp[it\mathbf{H}]\ket{0}^{\otimes n}$ and $\ket{\psi_{\rm loc}}:=\exp[itH_{\rm loc}]\ket{0}^{\otimes n}$. The sampled infidelity increases with $\varepsilon$ and exhibits quadratic scaling at small $\varepsilon$, consistent with an $O(\varepsilon)$ state perturbation. Because a new $V$ is drawn at each point, this sweep does not establish monotonicity for a fixed perturbation.

\subsection{Scope comparisons}\label{ssec:auxiliary}

The following comparisons test the short-time Choi and state-space identities and examine information beyond the scalar $\rhoH$. Cutting overheads depend on the target gate and protocol; Trotter error depends on the commutator $[\Hsep,\Hns]$; and finite-time entanglement and long-time operator growth depend on further dynamical structure.

\paragraph{Choi tripartite-information coefficient.} Proposition~\ref{prop:choi-exact} predicts $c(t)=2\ln(1/t)+a+o(1)$ with $a=1+H_{\mathrm{Sh}}(p)+\ln(d/\|\Hns\|_\HS^2)$ for the normalised ratio $c(t)=-I_3(A{:}C{:}D)/[t^2\rhoH\,d^{-1}\|\mathbf{H}\|_\HS^2]$ (Choi-leg labels as in \S\ref{ssec:connections}). Table~\ref{tab:exp10} tests this on six random two-local instances per partition. A two-parameter fit $b\ln(1/t)+a_{\rm fit}$ over the coarse grid $t\in\{0.005,0.01,0.02\}$ recovers the logarithmic law (root-mean-square residual at most $0.008$, against about $0.2$ for a linear-in-$t$ law) with $b$ within $2.5\%$ of $2$, but the grid is not asymptotic ($t\|\mathbf{H}\|_\op$ is not small at the larger $d$), which biases $b$ upward and $a_{\rm fit}$ downward by $0.03$--$0.24$. Evaluating $c(t)-2\ln(1/t)$ directly at $t=10^{-4}$ reproduces the predicted $a$ to four decimal places on every one of the $36$ instances. The offset varies from instance to instance through $H_{\mathrm{Sh}}(p)$ and $\|\Hns\|_\HS$ (standard deviations $0.14$--$0.50$), and its downward drift with $d$ in this table reflects the ensemble (unnormalised Gaussian coefficients make $\|\Hns\|_\HS^2/d$ grow with the number of cross-cut strings), not a universal dependence on dimension alone.

\begin{table}[!htbp]\centering\small
\begin{tabular}{ccccc}
\toprule
$(d_{S_1}, d_{S_2})$ & $d$ & $b$ (fit) & $a_{\rm fit}$ & $a$ (Proposition~\ref{prop:choi-exact}) \\
\midrule
$(2,2)$ & $4$  & $2.006\pm0.003$ & $-0.20\pm0.50$ & $-0.17\pm0.50$ \\
$(2,4)$ & $8$  & $2.023\pm0.007$ & $-1.25\pm0.31$ & $-1.13\pm0.28$ \\
$(4,4)$ & $16$ & $2.032\pm0.005$ & $-1.64\pm0.28$ & $-1.47\pm0.26$ \\
$(2,8)$ & $16$ & $2.030\pm0.006$ & $-1.42\pm0.30$ & $-1.25\pm0.27$ \\
$(4,8)$ & $32$ & $2.039\pm0.004$ & $-1.66\pm0.15$ & $-1.45\pm0.14$ \\
$(8,8)$ & $64$ & $2.044\pm0.008$ & $-1.81\pm0.28$ & $-1.58\pm0.24$ \\
\bottomrule
\end{tabular}
\caption{Check of Proposition~\ref{prop:choi-exact} on six random two-local Hamiltonian instances per partition (entropies in nats; mean $\pm$ standard deviation, ddof $=1$, over instances). Columns $3$--$4$: two-parameter fit $c(t)\approx b\ln(1/t)+a_{\rm fit}$ over the coarse grid $t\in\{0.005,0.01,0.02\}$. Column $5$: the predicted offset $a=1+H_{\mathrm{Sh}}(p)+\ln(d/\|\Hns\|_\HS^2)$ computed from each instance's $\Hns$; the direct value $c(10^{-4})-2\ln 10^{4}$ agrees with it to four decimal places on every instance, so the residual gap to $a_{\rm fit}$ is finite-$t$ bias of the coarse grid. The instance-mean $c(t)$ at the three grid points, for example $10.43/9.04/7.65$ for $(2,2)$, falls by $2\ln2\approx1.39$ per doubling of $t$.}\label{tab:exp10}
\end{table}

\paragraph{State-space entangling-power coefficient.} Proposition~\ref{prop:ep-exact} was checked by Monte Carlo ($2\times10^4$ Haar-random product states per instance) at $t=10^{-3}$ on random two-local instances with $(d_A,d_B)\in\{(2,2),(2,4),(4,4)\}$: the ratio of the measured $e_p(\exp[it\mathbf{H}])$ to $2t^2\|\Hns\|_\HS^2/((d_A+1)(d_B+1))$ lies in $[0.993,1.007]$ on all six instances tested, consistent with the $O(t^3)$ remainder and the sampling error.

\paragraph{Canonical-gate overheads.} Table~\ref{tab:exp5} places exact product-unitarity beside gate-cutting sampling overheads $\gamma^2$. The single-gate column is the optimal quasi-probability overhead for cutting one instance of the gate, $\gamma=1+2\Delta_U$ with $\Delta_U=\sum_{\alpha\ne\beta}|u_\alpha||u_\beta|$ in terms of the canonical (Cartan) coefficients of $U\simeq\sum_\alpha u_\alpha\,\sigma_\alpha\otimes\sigma_\alpha$~\cite[Cor.~3.1]{Schmitt2025}; it coincides with the Mitarai--Fujii value~\cite[Eq.~(20)]{Mitarai2021a} where their formula is optimal, and with the values of Piveteau--Sutter~\cite[Thm~4.3, Cor.~4.4]{Piveteau2023} for Clifford, controlled-rotation and Pauli-rotation gates. Classical communication gives no advantage for a single instance~\cite{Piveteau2023,Schmitt2025}. The joint-cut column is the asymptotic per-gate overhead $(1+\Delta_U)^2$ when many copies are cut together~\cite[Eq.~(6)]{Schmitt2025}, which for Clifford gates reproduces the $O(4^k)$ (CNOT) and $O(16^k)$ (SWAP) rates of~\cite{Piveteau2023}. The $\rhoH$ column is a principal-log diagnostic for the chosen representative $\mathbf{H}=-i\log U$ and is therefore sensitive to logarithm-branch and global-phase conventions; the ``Exact-sep?'' column records whether the unitary itself factors as $U_1\otimes U_2$ (operator-Schmidt rank $1$). At the boundary, the tabulated product gates have $\rhoH=0$ and cutting overhead $1$ (a product gate can have nonzero principal-log $\rhoH$; see the caption), and every non-product gate must be priced by a cutting protocol to run across the partition. SWAP is non-product across the two-qubit partition although it creates no entanglement from product inputs.

\begin{table}[!htbp]\centering\small
\begin{tabular}{lccccc}
\toprule
Gate & Exact-sep? & $\rhoH$ & $\mathrm{osr}_\cS(U)$ & single-gate $\gamma^2$ & joint-cut $\gamma^2$ per gate \\
\midrule
$I\otimes I$ & yes & $0.000$ & $1$ & $1$ & $1$ \\
$X\otimes I$ & yes & $0.000$ & $1$ & $1$ & $1$ \\
$\mathrm{CNOT}$ & no & $0.250$ & $2$ & $9$ & $4$ \\
$\mathrm{CZ}$ & no & $0.250$ & $2$ & $9$ & $4$ \\
$\mathrm{iSWAP}$ & no & $1.000$ & $4$ & $49$ & $16$ \\
$\mathrm{SWAP}$ & no & $0.750$ & $4$ & $49$ & $16$ \\
$\sqrt{\mathrm{SWAP}}$ & no & $0.750$ & $4$ & $34.27$ & $11.74$ \\
$\mathrm{CP}(0.785)$ & no & $0.250$ & $2$ & $3.115$ & $1.911$ \\
$\mathrm{CP}(1.57)$ & no & $0.250$ & $2$ & $5.826$ & $2.913$ \\
$\mathrm{CP}(3.14)$ & no & $0.250$ & $2$ & $9.000$ & $4.000$ \\
$\exp[i0.785\,X\!\otimes\! X]$ & no & $1.000$ & $2$ & $9.000$ & $4.000$ \\
$\exp[i1.57\,X\!\otimes\! X]$ & no & $1.000$ & $2$ & $1.006$ & $1.003$ \\
\bottomrule
\end{tabular}
\par\smallskip
\begin{minipage}{0.97\textwidth}
\raggedright\footnotesize
Single-gate $\gamma^2=(1+2\Delta_U)^2$: optimal sampling overhead for cutting one instance of the gate~\cite{Schmitt2025}; joint-cut $\gamma^2$ per gate $=(1+\Delta_U)^2$: asymptotic per-gate overhead when many copies are cut jointly~\cite{Schmitt2025}; CNOT = controlled-NOT; CZ = controlled-Z; CP = controlled phase. Eigenphases of $-i\log U$ are taken in $(-\pi,\pi]$, with the boundary eigenvalue $-1$ assigned phase $+\pi$. $\Delta_U$ is computed from the operator-Schmidt coefficients $s_\alpha=2|u_\alpha|$ of $U$ across the cut as $\Delta_U=[(\sum_\alpha s_\alpha)^2-4]/4$. For $\mathrm{CP}(\theta)$, $\Delta_U=|\sin(\theta/2)|$, so the single-gate overhead is $(1+2|\sin(\theta/2)|)^2$, equal to $9$ at $\theta=\pi$, consistent with $\mathrm{CP}(\pi)=\mathrm{CZ}$; for $\exp[i\theta X\otimes X]$, $\Delta_U=|\sin 2\theta|$, giving $9$ at the CNOT-equivalent angle $\theta=\pi/4$ and approaching $1$ at the near-product angle $\theta=1.57$.
\end{minipage}
\caption{Product-unitary separability, principal-log $\rhoH$, operator-Schmidt rank of the unitary $U$, and quasi-probabilistic cutting overheads for canonical two-qubit gates. Here $\rhoH$ is a generator-level diagnostic of $\mathbf{H}=-i\log U$ (principal matrix logarithm), whereas $\mathrm{osr}_\cS(U)$ is the operator-Schmidt rank of the unitary $U$ itself (for the two-qubit partition); the two columns describe different objects. The nonzero $\rhoH$ magnitudes depend on the logarithm-branch and global-phase conventions and are not gate invariants. For the gates listed, $\rhoH=0$ coincides with the ``Exact-sep?'' column, but in general only the implication $\rhoH(-i\log U)=0\Rightarrow U$ product holds at fixed branch: a product gate can have nonzero principal-log $\rhoH$ because the generator of a product unitary need not be block-additive (e.g.\ $\exp[i\tfrac{\pi}{2}Z\otimes Z]=Z\otimes(iZ)$ is a product with principal-log $\rhoH=1$, the single-time resonance of Theorem~\ref{thm:bipartite-single-t}; the tabulated $\exp[i\theta X\otimes X]$ angles sit off this resonance, so their ``no'' entries are exact). Both overhead columns are taken from the cited cutting literature.}\label{tab:exp5}
\end{table}

\begin{figure}[!htbp]\centering
\figIfExists[width=150mm]{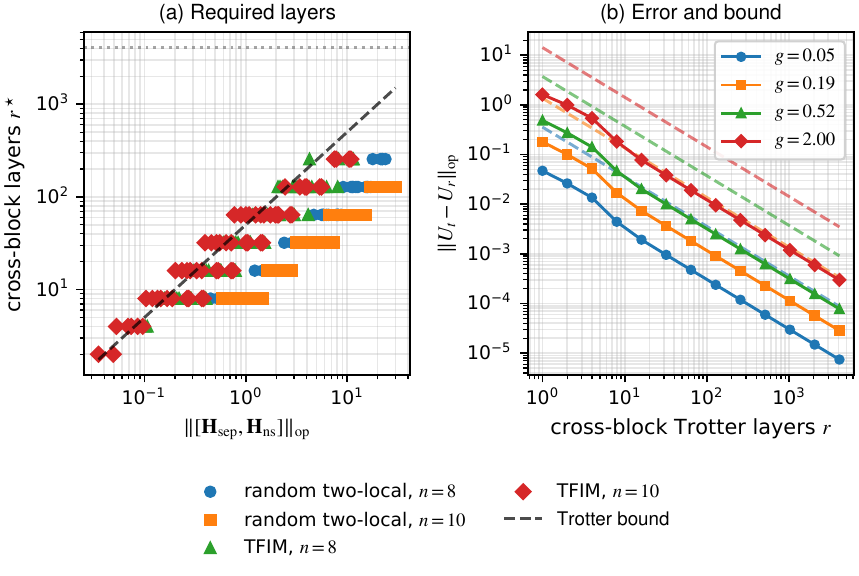}
\caption{First-order Trotterisation along the module partition, $U_r=(e^{it\Hsep/r}e^{it\Hns/r})^r$ at $t=1$ ($n\in\{8,10\}$, balanced cut, five instances per family, cross-block strength $g\in[0.05,2]$). Panel (a): cross-block layers $r^\star$ needed for $\|U_t-U_r\|_\op\le10^{-2}$ against $\|[\Hsep,\Hns]\|_\op$, pooled over instances and $g$; the dashed line is the product-formula bound $r=t^2\|[\Hsep,\Hns]\|_\op/(2\varepsilon)$; because $r$ is sampled at powers of two, points lie at or below the first sampled power of two above the line, and the transverse-field Ising family (one cut bond) nearly attains it. Panel (b): $\|U_t-U_r\|_\op$ against $r$ for one random two-local instance at $n=10$ and four values of $g$, with the bound $t^2\|[\Hsep,\Hns]\|_\op/(2r)$ dashed.}\label{fig:exp15}
\end{figure}

\paragraph{Trotterisation along the partition.} The partition-aware use of a product formula splits the generator as $\mathbf{H}=\Hsep+\Hns$: the block-additive part is run by the modules themselves, and the cross-block part is paid for in $r$ layers,
\[
U_r:=\big(e^{it\Hsep/r}\,e^{it\Hns/r}\big)^r\quad(r\ge1),\qquad U_0:=e^{it\Hsep},
\]
so that $U_0$ is the network-free approximation of Theorem~\ref{thm:approx} and each further layer costs one round of cross-module operations. The standard first-order product-formula bound~\cite{Childs2021} (operator-scrambling analyses of Trotter error such as~\cite{Feng2025} refine it for specific observables) is controlled by the same commutator that enters Theorem~\ref{thm:schmidt-refinement}, $\|U_t-U_r\|_\op\le t^2\|[\Hsep,\Hns]\|_\op/(2r)$, so the number of cross-block layers $r^\star(\varepsilon)$ needed for a tolerance $\varepsilon$ should scale with $\|[\Hsep,\Hns]\|_\op$, which is linear in the cross-coupling strength, whereas $\rhoH$ is quadratic in it at small coupling. Figure~\ref{fig:exp15} tests this on two families with a balanced cut, $n\in\{8,10\}$ and five instances each, random two-local Hamiltonians and open TFIM chains, in both cases with the cross-block part rescaled, $\mathbf{H}(g)=\Hsep+g\,\Hns/\|\Hns\|_\op$ for twelve values $g\in[0.05,2]$, at $t=1$ with $r\le4096$; no point is censored. Within each family and size, $r^\star$ grows with the commutator norm with log-log slopes $0.80$--$1.05$ for $\varepsilon\le10^{-2}$, and every point lies at or below the smallest sampled power of two above the bound $r=t^2\|[\Hsep,\Hns]\|_\op/(2\varepsilon)$ (that is, $r^\star<2\times$ the bound; the error itself never exceeds the bound). For the TFIM family, whose $\Hns$ is a single bond, the bound is nearly attained ($r^\star$ at $0.7$--$0.9$ of the bound in median, and $\|U_t-U_r\|_\op$ at $0.57$ of $t^2\|[\Hsep,\Hns]\|_\op/(2r)$ for $r\ge256$), while for random two-local instances the error stays $4$--$8$ times below the bound (at $0.11$ of it for $r\ge256$). Over the same sweep $\rhoH$ spans more than five orders of magnitude and does not predict $r^\star$ on its own: at equal $g$ the TFIM instances have $\rhoH$ one to two orders of magnitude larger than the random ones yet need about the same number of layers, the smaller TFIM commutator being offset by the random family's slack below the bound. The layer-free approximation $r=0$ reproduces Theorem~\ref{thm:approx}: $\|U_t-U_0\|_\op/(|t|\,\|\Hns\|_\op)$ is $0.996$ in median for the TFIM family, where a single cut bond makes the linear bound nearly tight, and $0.39$ for the random two-local instances.

\begin{figure}[!htbp]\centering
\figIfExists[width=150mm]{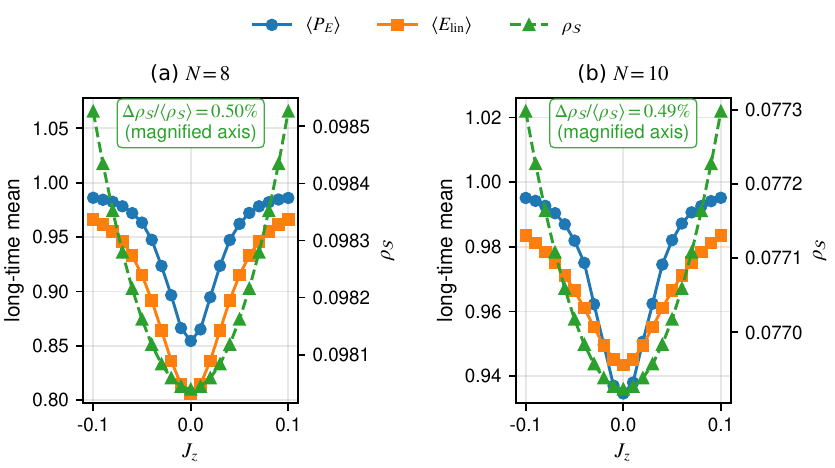}
\caption{Dip near integrability for the XYZ chain with a longitudinal field ($J_x=0.75$, $J_y=0.25$, $h=0.5$), shown for two sizes $N=8$ (a) and $N=10$ (b). The long-time Pauli entangling power $P_E$ (operator entanglement averaged over sampled non-identity Pauli strings) and the operator entanglement $E_{\rm lin}(U_t)$ of the bare evolution unitary (left axis; both averaged over $t\in\{15,17.5,20,22.5,25\}$) dip at $J_z=0$, while the generator-level $\rhoH$ (right, magnified axis) is the closed-form rational function of the text, nearly flat over $J_z\in[-0.1,0.1]$ (varying by under $1\%$) and even in $J_z$.}\label{fig:exp8}
\end{figure}

\paragraph{Integrability is invisible to $\rhoH$.} Figure~\ref{fig:exp8} shows that $\rhoH$ does not detect integrability. It revisits the integrability point studied in~\cite{Andreadakis2026} on the one-dimensional XYZ spin chain with a longitudinal $Z$ field,
\[
\mathbf{H}=\sum_i (J_x X_iX_{i+1}+J_y Y_iY_{i+1}+J_z Z_iZ_{i+1})+h\sum_i Z_i,
\]
with $J_x=0.75$, $J_y=0.25$, $h=0.5$, $J_z\in[-0.1,0.1]$, and open (rather than the reference's periodic) boundary conditions. On the half-chain cut only the one bond crossing the cut is non-separable, so $\rhoH=(J_x^2+J_y^2+J_z^2)/[(N-1)(J_x^2+J_y^2+J_z^2)+Nh^2]$ in closed form: a rational, even function of $J_z$, stationary at $J_z=0$ and varying by $0.5\%$ over the window. The plotted Pauli entangling power is an imported long-time operator-growth diagnostic: it estimates $P_E(U_t)=\mathbb{E}_{P\ne I}\,E_{\rm lin}(U_t^\dagger PU_t)$ (Heisenberg picture), where $E_{\rm lin}$ is the same linear operator entanglement as in \S\ref{ssec:connections}, here taken across the half-chain cut (sites $1,\dots,\lfloor N/2\rfloor$ versus the rest), and the expectation is over $P$ drawn uniformly from the $4^N-1$ non-identity Pauli strings ($100$ samples at each of $t\in\{15,17.5,20,22.5,25\}$; Appendix~\ref{app:repro}). It measures long-time spreading of initially factorised Pauli operators. At the integrable point $J_z=0$, this long-time diagnostic dips while $\rhoH$ stays flat to within $0.5\%$. Thus $\rhoH$ captures the short-time non-separable mass fraction but not long-time integrability, which depends on spectral and conservation-law structure.

\begin{figure}[!htbp]\centering
\figIfExists[width=110mm]{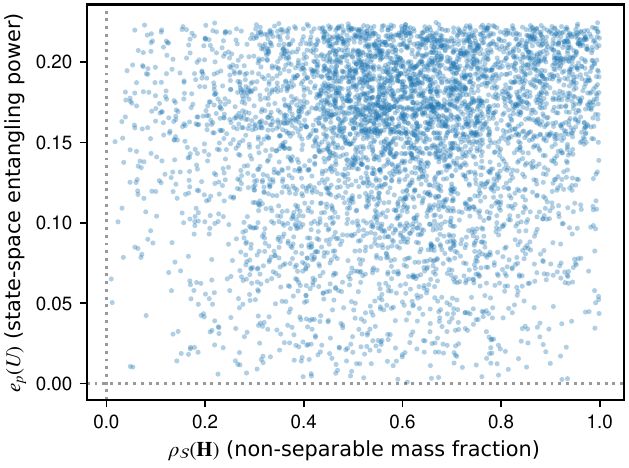}
\caption{State-space entangling power $e_p(U)$ vs.\ $\rhoH(\mathbf{H})$ of the sampling generator for $5000$ two-qubit gates $U=\exp[i\mathbf{H}]$ generated by random Cartan interaction terms $(c_x,c_y,c_z)$ and local $Z$ fields $(h_1,h_2)$, all uniform on $[0,\pi)$. The two are nearly uncorrelated (Pearson correlation $\approx0.1$).}\label{fig:exp4}
\end{figure}

\paragraph{State-space entangling-power scatter.} The state-space entangling power $e_p$ is the average entanglement a bipartite unitary (here two-qubit) creates from product inputs, a unitary-level finite-time quantity, whereas $\rhoH$ is a generator coefficient. Figure~\ref{fig:exp4} plots $e_p(U)$ against $\rhoH(\mathbf{H})$ of the sampling generator (not the principal-log value of Table~\ref{tab:exp5}) for $5000$ two-qubit gates $U=\exp[i\mathbf{H}]$ with $\mathbf{H}=c_xXX+c_yYY+c_zZZ+h_1\,Z\!\otimes\!I+h_2\,I\!\otimes\!Z$, all five coefficients drawn uniformly from $[0,\pi)$: Cartan interaction terms with random local $Z$ fields. The local fields are what move $\rhoH$ away from $1$ (a bare three-parameter Cartan generator has $\rhoH=1$ whenever it is nonzero, since every term is cross-block). The two quantities are nearly uncorrelated (Pearson correlation $\approx0.1$). The exact identity~\eqref{eq:ep-exact} of Proposition~\ref{prop:ep-exact} governs the flow as $t\to0$, where $e_p$ is set by the mass $\|\Hns\|_\HS^2$; the sampled evolutions are evaluated at $t=1$, outside that regime.

\medskip\noindent Together, the experiments support the short-time coefficients and approximation bounds, while showing why $\rhoH$ alone does not predict finite-time entanglement, integrability, or protocol cost.

\section{Related work}\label{sec:related}

The fixed-partition projection connects the present certificate to circuit cutting, operator-entanglement diagnostics, quantum mereology, and scrambling resource theory. In particular, its residual norm already appears in the ADZ and ZDA scrambling rates; here the same projection supplies an exact all-time product criterion and bounds the error of block-parallel evolution.

\paragraph{Circuit cutting.}
All of these methods begin from a nonlocal operation and minimise the cost of realising it on smaller devices. Peng, Harrow, Ozols, and Wu~\cite{Peng2020} cut wires and insert Pauli measurement/preparation operations, at exponential classical recombination cost in the number of cuts; gate-cut decompositions replace a nonlocal gate by quasi-probabilistic mixtures of local channels~\cite{Mitarai2021,Mitarai2021a}; CutQC~\cite{Tang2021} automates cut placement and reconstruction; and classical communication lowers the overhead within limits made precise by Piveteau--Sutter~\cite{Piveteau2023} and Brenner--Piveteau--Sutter~\cite{Brenner2025}. More recent variants address joint cutting of several two-qubit gates~\cite{Schmitt2025}, multi-controlled-gate cuts via the ZX calculus~\cite{Ufrecht2023}, QAOA workloads~\cite{Bechtold2023}, randomised-measurement cutting~\cite{Lowe2023}, wire--gate bridging in the ZX calculus~\cite{Schumann2025}, and noise-aware selection of cutting strategies~\cite{Pal2026}; entanglement-assisted quantum networking is surveyed in~\cite{Li2023}, and distributed-execution work such as~\cite{Sundaram2023} distributes a circuit over a network so as to minimise the number of teleportations. Our certificate asks the prior question, namely whether the target Hamiltonian evolution is already product across the fixed partition, with Theorem~\ref{thm:approx} quantifying the projection error when it is not.

\paragraph{Operator entanglement.}
The canonical two-qubit decomposition of Kraus and Cirac~\cite{Kraus2001} isolates the entangling part of a unitary into three angles (the Cartan form), while product-structure methods recognise tensor structure from unitary access~\cite{Mansuroglu2024}. The state-space entangling power of Zanardi--Zalka--Faoro~\cite{Zanardi2000}, operator entanglement~\cite{Zanardi2001}, and the ADZ operator-space entangling power~\cite{Andreadakis2024} quantify different forms of entanglement generation. Hosur et al.~\cite{Hosur2016} give the complementary Choi-state formulation through tripartite information. Propositions~\ref{prop:ep-exact} and~\ref{prop:rho-Ep} express the two entangling-power coefficients in terms of $\|\Hns\|_\HS^2$, and Proposition~\ref{prop:choi-exact} derives the leading Choi-information asymptotics from the Schmidt spectrum of $\Hns$.

\paragraph{Quantum mereology.}
Quantum mereology asks which tensor-product structure (TPS) should be assigned to a quantum system rather than assuming the factorisation in advance. Soulas~\cite{Soulas2025} asks whether some TPS keeps a given time-evolving state unentangled; related work on preferred structures and spectral tests develops the same variable-subsystem viewpoint~\cite{Soulas2025a,Loizeau2025}, and the minimal-scrambling programme of~\cite{Zanardi2024} selects subsystems by minimising the Gaussian scrambling rate. ZDA~\cite{Zanardi2026} go further by varying a family of generalised TPSs and using an algebraic scrambling functional to detect mereological phase transitions. Here the partition $\cS$ is fixed by a hardware layout, modular architecture, or compiler assignment; Proposition~\ref{prop:zda-identification} translates between the two views, since the ZDA projection off the fixed factor algebra plus commutant is exactly our projection onto $\Hns$, with squared scrambling functional $\|\Hns\|_\HS^2=\rhoH(\mathbf{H})\|\mathbf{H}\|_\HS^2$.

\paragraph{Scrambling resources.}
Garcia, Bu, and Jaffe~\cite{Garcia2023} formulate a resource theory of scrambling for unitaries, with entanglement-scrambling and magic-scrambling layers based on Pauli growth and out-of-time-ordered-correlator (OTOC) magic. Andreadakis--Zanardi~\cite{Andreadakis2026} prove that a unitary $U$ leaves all Heisenberg-evolved Pauli strings unentangled across a bipartition exactly when $U^\dagger=(V_A\otimes W_B)\,C$ with $C$ Clifford and $V_A,W_B$ local, tying the operator entanglement that underlies $\rhoH$ to nonlocal nonstabilizerness generation. These theories are unitary-level: the Pauli-growth monotone of~\cite{Garcia2023} uses single-site Pauli weight rather than a supplied module partition, and~\cite{Andreadakis2026} works across a given bipartition but does not treat the partition as Hamiltonian-level data. The present construction keeps the partition as data. Within-block operator growth is treated as free, while cross-block growth contributes, at leading short-time order, exactly through $\Hns$, whose ADZ/ZDA scrambling rate is thereby tied to the all-time separability boundary.

Table~\ref{tab:related} compares the inputs and outputs of representative methods. The present certificate addresses product evolution before a cutting or decomposition protocol is selected.
\begin{table}[!htbp]\centering\small
\begin{tabular}{@{}L{0.18\textwidth}L{0.31\textwidth}L{0.20\textwidth}L{0.22\textwidth}@{}}
\toprule
Work & Question answered & Object & Output \\
\midrule
Peng et al.~\cite{Peng2020} & How can a circuit be evaluated after cutting wires? & Circuit with wire cuts & Local experiments plus quasi-probabilistic recombination \\
Mitarai--Fujii~\cite{Mitarai2021,Mitarai2021a} & How can a nonlocal gate be replaced by local operations? & Gate cut & Sampling overhead for local-channel decomposition \\
Piveteau--Sutter~\cite{Piveteau2023}; Brenner--Piveteau--Sutter~\cite{Brenner2025} & How much can classical communication reduce cutting overhead? & Cut gates/circuits & Assisted cutting decompositions and overhead trade-offs \\
CutQC~\cite{Tang2021} & Where should a large circuit be cut and reconstructed? & Full circuit workflow & Automated cut placement and reconstruction \\
Mansuroglu et al.~\cite{Mansuroglu2024} & Can a unitary's tensor-product decomposition be extracted from Choi-state data? & Unitary access with Choi-state tomography & Tensor-product decomposition of the accessed unitary \\
This work & Does this Hamiltonian evolution factor on the fixed partition, and what is the projection error otherwise? & Supplied Hamiltonian $\mathbf{H}$ and partition $\cS$ & $\rhoH=0$ certifies product evolution; otherwise error bounds and rank diagnostics \\
\bottomrule
\end{tabular}
\caption{Problem-level positioning against representative cutting and decomposition methods. Cutting methods price or organise nonlocal operations; the present certificate is Hamiltonian-level and identifies when the fixed partition has no cross-block interaction to cut, and bounds or estimates the projection error when it does.}\label{tab:related}
\end{table}

\section{Conclusion and outlook}\label{sec:conclusion}

\subsection{Distributed-computing consequence}\label{ssec:distributed-consequence}

For a supplied Hamiltonian and a fixed module partition, the projection residual $\Hns$ gives a direct test of exact independent execution: the evolution is product for all times if and only if $\Hns=0$, equivalently $\rhoH=0$. For Pauli-list inputs, the test requires one pass over the combined coefficients. When the residual is nonzero, omitting it incurs operator-norm error at most $|t|\,\|\Hns\|_\op$. For a bipartition, the first-order correction has the same operator-Schmidt rank as $\Hns$ and leaves a quadratic residual, providing a structured approximation whose implementation cost depends on the chosen protocol.

Across a bipartition, $\|\Hns\|_\HS^2$ determines the leading state-space and operator-space entangling-power coefficients and equals the squared fixed-g-TPS ZDA scrambling functional. The Schmidt spectrum of $\Hns$ also determines the leading Choi tripartite-information expansion. Numerical checks support these identities and bounds, and the block-factorised benchmark demonstrates exact evolution up to $20$ qubits. The certificate is therefore useful before compilation, while finite-time entanglement, long-time dynamics, and cutting costs require information beyond $\rhoH$.

\subsection{Outlook}\label{ssec:outlook}

\emph{Higher-order corrections.} Extending Theorem~\ref{thm:schmidt-refinement} requires tracking the product-term structure of higher-order terms and nested commutators. This would relate correction rank to accuracy beyond first order. Higher-order Choi asymptotics could likewise quantify the finite-time departures observed in Table~\ref{tab:exp10}.

\emph{Local-observable bounds.} For finite-range, bounded interactions on a lattice, Lieb--Robinson estimates~\cite{Lieb1972} could sharpen the global error bound for observables far from the partition boundary. The resulting guarantee would depend on the distance to omitted interactions as well as the evolution time.

\emph{Hardware and learning interfaces.} Comparing monolithic and block-parallel evolution on the same hardware instances would place the projection error alongside the sampling and noise costs of cutting. When the Hamiltonian is unknown, estimating its cross-block coefficients through Hamiltonian learning, potentially using classical-shadow measurements~\cite{Huang2020}, would require confidence intervals on the residual mass.

\emph{Partition selection.} For sparse Pauli inputs, minimising $\sum_{P:\,w_\cS(P)\ge2}c_P^2=d^{-1}\|\Hns\|_\HS^2$ subject to module capacities and connectivity constraints gives a weighted hypergraph-partitioning problem: qubits are vertices and Pauli supports are hyperedges weighted by $c_P^2$. This would adapt the hypergraph approach used in distributed-circuit compilation~\cite{Andres2019} to a Hamiltonian-level objective.

\appendix
\section{Reproducibility details}\label{app:repro}

The code and data archive at \href{https://doi.org/10.5281/zenodo.23073840}{doi:10.5281/zenodo.23073840} contains the experiment scripts, fixed random seeds, sampling grids and software versions, with figure and table data and timing controls in \texttt{data/}. Timings are medians of five repetitions on one NVIDIA Grace Hopper node at Pawsey Ella. The sparse baseline and tensor contractions run on the CPU; block eigendecompositions use the GPU for block dimension $\ge64$. The dense baseline is omitted when its estimated storage exceeds $240$~GiB. CPU-only controls, reported in \S\ref{ssec:exp3}, distinguish the algorithmic gain from GPU acceleration.

\section{Detailed proofs}\label{app:proofs}

All spaces in this appendix are finite-dimensional. Except in Theorem~\ref{thm:schmidt-refinement}, which permits any self-adjoint additive splitting, $\Hsep$ and $\Hns$ are the orthogonal projection pair of~\eqref{eq:Hsep-closed}. In the short-time expansions, the Hamiltonian and partition are fixed; remainder constants may depend on them and on the block dimensions. An $O(t^k)$ remainder is bounded in magnitude by a constant times $|t|^k$.

\subsection{Proof of Proposition~\ref{prop:HS}}\label{app:HS}
\begin{proof}[Proof of Proposition~\ref{prop:HS}]
Use the same Duhamel identity as in Theorem~\ref{thm:approx}:
\[
\exp[itA]-\exp[itB]
=i\int_0^t\exp[i(t-s)A](A-B)\exp[isB]\,ds .
\]
Set $A=\mathbf{H}$ and $B=\Hsep$. The Hilbert--Schmidt norm is invariant under left and right multiplication by unitaries, so the integrand has Hilbert--Schmidt norm exactly $\|\Hns\|_\HS$ for every $s$. Using $\big\|\int_0^t g(s)\,ds\big\|_\HS\le|t|\sup_s\|g(s)\|_\HS$ for both signs of $t$,
\[
\|\exp[it\mathbf{H}]-\exp[it\Hsep]\|_\HS
\le |t|\,\|\Hns\|_\HS .
\]
Dividing by $\sqrt d$ gives the unitary-normalised form; for qubits, Pauli orthogonality gives $\|\Hns\|_\HS^2/d=\sum_{w_\cS(P)\ge2}c_P^2$. For $\mathbf{H}\ne0$, dividing instead by $\|\mathbf{H}\|_\HS$ and using
$\rhoH(\mathbf{H})=\|\Hns\|_\HS^2/\|\mathbf{H}\|_\HS^2$ gives the displayed normalised bound.
\end{proof}

\subsection{Proof of Theorem~\ref{thm:schmidt-refinement}}\label{app:schmidt-refinement}
\begin{proof}[Proof of Theorem~\ref{thm:schmidt-refinement}]
\emph{Step 1 (Duhamel).} For Hermitian $\mathbf{H}$ and $\Hsep$,
\begin{equation}\label{eq:duhamel-restated}
\exp[it\mathbf{H}]-\exp[it\Hsep] \;=\; i\int_{0}^{t}\exp[i(t-s)\mathbf{H}]\,\Hns\,\exp[is\Hsep]\,ds .
\end{equation}
Set $f(s):=\exp[i(t-s)\mathbf{H}]\,\Hns\,\exp[is\Hsep]$.

\emph{Step 2 (Integration by parts to extract the right-endpoint term).} Differentiating $s\mapsto s\,f(s)$ and integrating from $0$ to $t$,
\[
\int_0^t f(s)\,ds \;=\; t\,f(t) \;-\; \int_0^t s\,f'(s)\,ds .
\]
Multiplying by $i$ and using $f(t)=\Hns\,\exp[it\Hsep]$,
\begin{equation}\label{eq:RtIBP}
\exp[it\mathbf{H}]-\exp[it\Hsep] \;=\; \underbrace{it\,\Hns\,\exp[it\Hsep]}_{\,=\,T_t} \;-\; i\int_0^t s\,f'(s)\,ds .
\end{equation}

\emph{Step 3 (Differentiate $f$).} Using $\frac{d}{ds}\exp[i(t-s)\mathbf{H}]=-i\mathbf{H}\,\exp[i(t-s)\mathbf{H}]$ (which commutes with its generator) and $\frac{d}{ds}\exp[is\Hsep]=i\Hsep\,\exp[is\Hsep]$:
\[
f'(s) = \exp[i(t-s)\mathbf{H}]\,\big(-i\mathbf{H}\,\Hns + i\,\Hns\,\Hsep\big)\,\exp[is\Hsep].
\]
Substituting $\mathbf{H}=\Hsep+\Hns$ in the first term gives $-i\mathbf{H}\,\Hns + i\,\Hns\,\Hsep=-i\big([\Hsep,\Hns]+\Hns^2\big)$, so
\begin{equation}\label{eq:fprime}
f'(s) \;=\; -i\,\exp[i(t-s)\mathbf{H}]\,\big([\Hsep,\Hns]+\Hns^2\big)\,\exp[is\Hsep].
\end{equation}

\emph{Step 4 (Bound the residual).} The exponentials are unitary, so
\[
\|f'(s)\|_\op \le \kappa:=\|[\Hsep,\Hns]+\Hns^2\|_\op
\le 2\,\|\Hsep\|_\op\,\|\Hns\|_\op + \|\Hns\|_\op^2.
\]
Therefore, from~\eqref{eq:RtIBP},
\[
\|R_t\|_\op
= \Big\|i\int_0^t s\,f'(s)\,ds\Big\|_\op
\le \kappa\int_{\min(0,t)}^{\max(0,t)} |s|\,ds
=\frac{t^2}{2}\kappa,
\]
which is~\eqref{eq:Rt-comm-bound}. The triangle estimate displayed above then gives the coarser bound~\eqref{eq:Rt-bound}.

\emph{Step 5 (Schmidt rank).} If $\Hns=0$, then $r=0$ and $T_t=R_t=0$ for all $t$, so all conclusions hold. Otherwise take $t\ne0$. $\Hsep=H_1\otimes I+I\otimes H_2$ has commuting summands, so $\exp[it\Hsep]=\exp[itH_1]\otimes\exp[itH_2]$. Substituting~\eqref{eq:osr-def} into~\eqref{eq:Tt-def} gives
\[
\begin{aligned}
T_t &= it\sum_{k=1}^{r}\sigma_k\,(A_k\otimes B_k)\,(\exp[itH_1]\otimes\exp[itH_2])\\
&=it\sum_{k=1}^{r}\sigma_k\,(A_k\,\exp[itH_1])\otimes(B_k\,\exp[itH_2]).
\end{aligned}
\]
Right multiplication by a unitary is a Hilbert--Schmidt isometry on $\cB(\cH_{S_1})$ and on $\cB(\cH_{S_2})$ separately: for any unitary $V$ and any operators $M,N$, $\langle MV,NV\rangle_\HS=\tr(V^\dagger M^\dagger NV)=\tr(M^\dagger N)=\langle M,N\rangle_\HS$. Hence $\{A_k\exp[itH_1]\}_k\subset\cB(\cH_{S_1})$ and $\{B_k\exp[itH_2]\}_k\subset\cB(\cH_{S_2})$ remain Hilbert--Schmidt orthonormal; absorbing the unimodular scalar $i\,\mathrm{sign}(t)$ into the first factors (which preserves their orthonormality) exhibits the displayed expansion as an operator-Schmidt decomposition of $T_t$ with coefficients $|t|\sigma_k>0$. Therefore $\mathrm{osr}_\cS(T_t)=r$ for $t\ne0$, while $T_0=0$ gives $\mathrm{osr}_\cS(T_0)=0$.

\emph{Step 6 (First-order match).} The operator norm is right-invariant under unitary multiplication, so
\[
\|T_t\|_\op \;=\; \|it\,\Hns\,\exp[it\Hsep]\|_\op \;=\; |t|\,\|\Hns\|_\op .
\]
\end{proof}

\subsection{Proof of Proposition~\ref{prop:ep-exact}}\label{app:ep-exact}
\begin{proof}[Proof of Proposition~\ref{prop:ep-exact}]
\emph{Step 1 (remove local evolution).} Local output unitaries preserve $E_{\rm st}$, so $e_p(\exp[it\mathbf{H}])=e_p(M_t)$ for
\[
M_t:=\exp[-it\Hsep]\exp[it\mathbf{H}]=I+it\Hns+O(t^2).
\]
This expansion holds in operator norm, uniformly on unit input vectors; $M_t$ is exactly unitary.

\emph{Step 2 (linear entropy near a product state).} Fix unit vectors $\ket{\phi_1},\ket{\phi_2}$ and choose local bases with these as their zeroth vectors. Write $C_{ab}(t)$ for the coefficient matrix of $M_t\ket{\phi_1\phi_2}$ in these bases. It satisfies $\|C(t)\|_\HS=1$ and
\[
C_{ab}(t)=\delta_{a0}\delta_{b0}+t\chi_{ab}+O(t^2),
\qquad \ket\chi=i\Hns\ket{\phi_1\phi_2}.
\]
The identity expressing linear entropy through the $2\times2$ minors of a normalised coefficient matrix is
\[
1-\tr[(CC^\dagger)^2]
=2\sum_{a<a'}\sum_{b<b'}
\big|C_{ab}C_{a'b'}-C_{ab'}C_{a'b}\big|^2.
\]
It follows by expanding $(\tr CC^\dagger)^2-\tr[(CC^\dagger)^2]$ and grouping the terms with distinct row and column indices. The minors using row $0$ and column $0$ are $t\chi_{ab}+O(t^2)$ for $a,b>0$; every other minor is $O(t^2)$. Therefore, with $Q_i:=I-\ket{\phi_i}\bra{\phi_i}$,
\[
E_{\rm st}(M_t\ket{\phi_1\phi_2})
=2t^2\sum_{a,b>0}|\chi_{ab}|^2+O(|t|^3)
=2t^2\big\|(Q_1\otimes Q_2)\Hns\ket{\phi_1\phi_2}\big\|^2+O(|t|^3).
\]
Thus the unspecified second-order state correction cannot alter the $t^2$ coefficient. The operator-norm remainder in Step~1 and the finite minor sum give a bound uniform over unit product inputs, so the remainder may be Haar-averaged.

\emph{Step 3 (evaluate the Haar average).} For a Haar-random unit vector $\ket\phi\in\mathbb C^q$ and $Q=I-\ket\phi\bra\phi$, the second moment
$\mathbb E[(\ket\phi\bra\phi)^{\otimes2}]=(I+F)/(q(q+1))$, where $F$ swaps the two copies, gives, for arbitrary operators $X,Y$,
\[
\mathbb E\bra\phi X^\dagger QY\ket\phi
=\frac{\tr(X^\dagger Y)}{q+1}
-\frac{\overline{\tr X}\,\tr Y}{q(q+1)}.
\]
Expand $\Hns=\sum_{\mu,\nu}h_{\mu\nu}E_\mu\otimes F_\nu$ in traceless self-adjoint Hilbert--Schmidt-orthonormal local bases. The coefficients are real and $\sum_{\mu,\nu}h_{\mu\nu}^2=\|\Hns\|_\HS^2$. Independence of the two Haar averages and the displayed identity eliminate all cross terms, yielding
\[
\mathbb E\big\|(Q_1\otimes Q_2)\Hns\ket{\phi_1\phi_2}\big\|^2
=\frac{\sum_{\mu,\nu}h_{\mu\nu}^2}{(d_A+1)(d_B+1)}.
\]
Combining with Step~2 proves~\eqref{eq:ep-exact}. If either block is one-dimensional, or more generally $\Hns=0$, the evolution is local and $e_p$ vanishes identically; the same formula holds with zero coefficient.
\end{proof}

\subsection{Proof of Proposition~\ref{prop:choi-exact}}\label{app:choi-exact}
\begin{proof}[Proof of Proposition~\ref{prop:choi-exact}]
Throughout this proof $A,B,C,D$ denote the Choi legs, with $d_A=d_C=d_{S_1}$, $d_B=d_D=d_{S_2}$ and $d=d_{S_1}d_{S_2}$. We first take $t>0$ sufficiently small.

\emph{Step 1 (reduce to two marginal entropies).} The input and output marginals of a normalised unitary Choi state are $\varrho_{AB}=\varrho_{CD}=I/d$. Since the full state is pure, $S(\varrho_{ACD})=S(\varrho_B)=\ln d_B$. Thus $I(A{:}CD)=2\ln d_A$ and
\[
-I_3(A{:}C{:}D)=S(\varrho_{AC})+S(\varrho_{AD})-\ln d.
\]
Left multiplication of $U$ by $V_1\otimes V_2$ acts locally on the output legs $C,D$; right multiplication by $W_1\otimes W_2$ acts as $W_1^T\otimes W_2^T$ on the input legs $A,B$. Both marginal entropies are therefore invariant under either operation. Use the right-local transformation
\[
M_t:=U_t\exp[-it\Hsep]=I+it\Hns+R_t',
\qquad R_t':=R_t\exp[-it\Hsep].
\]
Theorem~\ref{thm:schmidt-refinement} gives
$\|R_t'\|_\op\le t^2\kappa/2$, where $\kappa=\|[\Hsep,\Hns]+\Hns^2\|_\op$. Also $M_t$ is unitary, so $\|M_t\|_\HS^2=d$.

\emph{Step 2 (the small Schmidt eigenvalues).} Reshaping $d^{-1/2}\mathrm{vec}(M_t)$ across $(A,C)\,|\,(B,D)$ shows that the nonzero eigenvalues of $\varrho_{AC}$ are $s_k(M_t)^2/d$, where $s_k$ are the operator-Schmidt coefficients. Choose
$\Hns=\sum_{k=1}^r\sigma_k A_k\otimes B_k$ as in~\eqref{eq:osr-def}. Both partial traces of $\Hns$ vanish. For example,
\[
A_k=\sigma_k^{-1}\tr_{S_2}[(I\otimes B_k^\dagger)\Hns],
\qquad
\tr A_k=\sigma_k^{-1}\tr[B_k^\dagger\tr_{S_1}\Hns]=0,
\]
and similarly $\tr B_k=0$. Hence
\[
I+it\Hns
=\sqrt d\,\frac{I_{S_1}}{\sqrt{d_{S_1}}}\otimes\frac{I_{S_2}}{\sqrt{d_{S_2}}}
+\sum_{k=1}^r(t\sigma_k)(iA_k)\otimes B_k
\]
is an operator-Schmidt decomposition. For sufficiently small $t$, its largest coefficient is $\sqrt d$ and the others are $t\sigma_1,\ldots,t\sigma_r$, followed by zeros. Reshaping preserves the Hilbert--Schmidt norm, and the singular-value perturbation bound~\cite{Mirsky1960} gives, for the singular values in decreasing order,
\[
|s_k(M_t)-s_k(I+it\Hns)|
\le\|R_t'\|_\HS\le\sqrt d\,\|R_t'\|_\op=O(t^2).
\]
No nondegeneracy of the $\sigma_k$ is needed. Label the largest eigenvalue of $\varrho_{AC}$ by $\lambda_0$ and pad its spectrum with zeros if necessary. Then
\[
\lambda_k=\frac{t^2\sigma_k^2}{d}+O(t^3)=\varepsilon_t p_k+O(t^3)\quad(1\le k\le r),
\qquad \lambda_k=O(t^4)\quad(k>r),
\]
and exact normalisation gives $\lambda_0=1-\sum_{k\ge1}\lambda_k$.

\emph{Step 3 (entropy of the rank-deficient marginal).} Set $\varepsilon:=\sum_{k\ge1}\lambda_k=\varepsilon_t+O(t^3)$. Each $p_k$ is positive and fixed, so for $1\le k\le r$ the eigenvalue $\lambda_k$ is bounded above and below by positive constants times $t^2$. The derivative of $x\ln x$ on these intervals is $O(\ln(1/t))$; hence replacing $\lambda_k$ by $\varepsilon_t p_k$ costs $O(t^3\ln(1/t))$. The remaining eigenvalues contribute $O(t^4\ln(1/t))$, using $0\ln0=0$. Consequently,
\begin{align*}
S(\varrho_{AC})
&=-(1-\varepsilon)\ln(1-\varepsilon)
  -\sum_{k=1}^r\lambda_k\ln\lambda_k+O(t^4\ln(1/t))\\
&=\varepsilon_t\big[1+\ln(1/\varepsilon_t)+H_{\mathrm{Sh}}(p)\big]
  +O(t^3\ln(1/t)).
\end{align*}
Here $\ln(1/\varepsilon_t)=2\ln(1/t)+\ln(d/\|\Hns\|_\HS^2)$. The estimate is for the fixed nonzero residual; it does not assert uniformity as its nonzero Schmidt coefficients tend to zero.

\emph{Step 4 (the full-rank $AD$ marginal).} Let $T_{S_1}$ denote partial transpose on the first physical block in the bases used for vectorisation. A direct contraction gives the exact identity
\[
\varrho_{AD}(M)=\frac1d M^{T_{S_1}}(M^{T_{S_1}})^\dagger,
\qquad
[\varrho_{AD}(M)]_{a\delta,a'\delta'}
=\frac1d\sum_{b,c}M_{c\delta,ab}\,\overline{M_{c\delta',a'b}}.
\]
Partial transpose preserves Hermiticity, so $\Hns^{T_{S_1}}$ is Hermitian. Applying Step~1 to this identity, the first-order terms cancel:
\[
\varrho_{AD}(M_t)
=\frac1d\big(I+it\Hns^{T_{S_1}}+O(t^2)\big)
             \big(I-it\Hns^{T_{S_1}}+O(t^2)\big)
=\frac Id+\Delta_t,
\]
where $\Delta_t$ is Hermitian, $\tr\Delta_t=0$ and $\|\Delta_t\|_\HS=O(t^2)$. For sufficiently small $t$ this marginal is positive definite. Taylor expansion about $I/d$ therefore yields
\[
S(\varrho_{AD})
=\ln d-\frac d2\|\Delta_t\|_\HS^2+O(\|\Delta_t\|_\HS^3)
=\ln d+O(t^4).
\]
Combining Steps~1, 3 and 4 proves~\eqref{eq:I3-leading}. Division by $\varepsilon_t>0$ gives the more explicit remainder
$c(t)=2\ln(1/t)+a+O(t\ln(1/t))$, which implies the stated $o(1)$.

Finally, the evenness assertion requires both marginals. The Choi vector of $U^\dagger$ is the complex conjugate of that of $U$ with input and output interchanged, $A\leftrightarrow C$ and $B\leftrightarrow D$. Thus $S(\varrho_{AC}(U^\dagger))=S(\varrho_{AC}(U))$, while $S(\varrho_{AD}(U^\dagger))=S(\varrho_{BC}(U))=S(\varrho_{AD}(U))$ by purity. Since $U_{-t}=U_t^\dagger$, $-I_3$ is even; the expansion for negative $t$ uses $|t|$ in the logarithms and remainder. If $\Hns=0$, the Choi state factors across $AC\,|\,BD$ and $-I_3=0$ for every $t$; the normalised ratio $c(t)$ and the distribution $p$ are not defined in that case.
\end{proof}

\subsection{Proof of Proposition~\ref{prop:zda-identification}}\label{app:zda-identification}
\begin{proof}[Proof of Proposition~\ref{prop:zda-identification}]
(i) The identity, the one-block traceless sectors and the two-block traceless sector are mutually Hilbert--Schmidt orthogonal. Complexifying the decomposition of \S\ref{sec:prelims} therefore gives, for every $X\in\cB(\cH)$,
\[
\Pi_{L_0}(X)
=\frac{\tr_{S_2}X}{d_{S_2}}\otimes I_{S_2}
+I_{S_1}\otimes\frac{\tr_{S_1}X}{d_{S_1}}
-\frac{\tr X}{d}I.
\]
The last term counts the common identity direction once. This projection commutes with the adjoint, and its restriction to self-adjoint operators is exactly the block-additive projection~\eqref{eq:Hsep-closed}. Hence $Q_0|_{\cB_s(\cH)}=\Pi_{\mathrm{ns}}$ and $Q_0(\mathbf{H})=\Hns$.

(ii) Substituting (i) into the defining functional gives
$\sigma_s^2(L_0,\mathbf{H})=\|Q_0(\mathbf{H})\|_\HS^2=\|\Hns\|_\HS^2$.
Equation~\eqref{eq:Hns-norm} supplies the expression in terms of $\rhoH$, including $\mathbf{H}=0$ under the stated convention. Multiplying the inner product by $1/d$ leaves the orthogonal projection unchanged and divides every squared norm in this identity by $d$.

(iii) We verify the short-time coefficient directly from the ZDA definition~\cite[Eq.~(3)]{Zanardi2026}:
\[
G_{L_0}(t)=\frac1{2d}\,\mathbb E_{X,Y}
\big\|[X,U_tYU_t^\dagger]\big\|_\HS^2,
\qquad X=V_1\otimes I_{S_2},\quad Y=I_{S_1}\otimes V_2,
\]
where $V_1,V_2$ are independent Haar-random local unitaries. Since $[X,Y]=0$ and $[X,[\Hsep,Y]]=0$, Taylor expansion gives
\[
[X,U_tYU_t^\dagger]=it[X,[\Hns,Y]]+O(t^2),
\]
uniformly in $X,Y$. For any $K\in\cB(\cH)$, Haar averaging over $V_1$ is the conditional expectation $\mathbb E_X X^\dagger KX=\mathcal E_2(K)$, the Hilbert--Schmidt projection onto the second-block algebra. Expanding the squared commutator gives
\[
\mathbb E_X\|[X,K]\|_\HS^2
=2\big(\|K\|_\HS^2-\|\mathcal E_2(K)\|_\HS^2\big).
\]
Both partial traces of $\Hns$ vanish, so
$\mathcal E_2([\Hns,Y])=[\mathcal E_2(\Hns),Y]=0$ and $\mathcal E_1(\Hns)=0$. Applying the same identity successively on the two blocks gives
\[
\mathbb E_{X,Y}\|[X,[\Hns,Y]]\|_\HS^2
=2\mathbb E_Y\|[\Hns,Y]\|_\HS^2
=4\|\Hns\|_\HS^2.
\]
Therefore
\[
G_{L_0}(t)=\frac{2t^2}{d}\|\Hns\|_\HS^2+O(|t|^3),
\]
and comparison with Proposition~\ref{prop:rho-Ep} proves $E_p^{\op}(\mathcal U_t)=2G_{L_0}(t)+O(|t|^3)$. With the normalised Hilbert--Schmidt norm, the explicit $1/d$ is absorbed into the squared residual norm, so the coefficients become $2$ and $4$ while both physical diagnostics and their ratio are unchanged. If either block is one-dimensional, or $\Hns=0$, the evolution is local and both diagnostics vanish identically.
\end{proof}


\ack{This work was carried out within the Pawsey Supercomputing Research Centre's Quantum Supercomputing Innovation Hub, made possible by a grant from the Australian Government through the National Collaborative Research Infrastructure Strategy (NCRIS). Computational resources were provided by the Pawsey Supercomputing Research Centre's Setonix Supercomputer (\href{https://doi.org/10.48569/18sb-8s43}{https://doi.org/10.48569/18sb-8s43}), with funding from the Australian Government and the Government of Western Australia. AI-assisted tools were used during the development of portions of the software implementation; all generated code and related documentation were reviewed, checked, and modified by the authors.}

\section*{Conflict of interest}
The authors declare no conflicts of interest.

\funding{National Natural Science Foundation of China, Grants 12271394 and 12071336. Resources were provided by the Pawsey Supercomputing Research Centre with funding from the Australian Government and the Government of Western Australia.}

\roles{K.~He and S.~Liu contributed equally to this work. All authors contributed to the conceptualisation, analysis, and writing of the manuscript and approved the final version.}

\data{The experiment code and numerical data are available on Zenodo at \href{https://doi.org/10.5281/zenodo.23073840}{doi:10.5281/zenodo.23073840}.}

\bibliographystyle{iopart-num}
\bibliography{ref}

\end{document}